\documentclass[sigconf]{acmart}
\usepackage{balance}
\usepackage{natbib}
\usepackage{array}
\usepackage{multirow}
\usepackage{xcolor}
\usepackage{algorithm}
\usepackage{algorithmic}
\usepackage{enumitem}

\usepackage[skip=2pt]{caption}

\AtBeginDocument{%
  \providecommand\BibTeX{{%
    \normalfont B\kern-0.5em{\scshape i\kern-0.25em b}\kern-0.8em\TeX}}}

\copyrightyear{2026}
\acmYear{2026}
\setcopyright{cc}
\setcctype{by-nc-nd}
\acmConference[UIST '26]{The 39th Annual ACM Symposium on User Interface Software and Technology}{November 02--05, 2026}{Detroit, MI, USA}
\acmBooktitle{The 39th Annual ACM Symposium on User Interface Software and Technology (UIST '26), November 02--05, 2026, Detroit, MI, USA}
\acmDOI{10.1145/3830398.3830598}
\acmISBN{979-8-4007-2856-3/2026/11}

\newcommand{\red}[1]{#1}

\begin{document}

\title{X-Hinges: 3D Printing Self-Sensing Compliant Mechanisms for Continuous and Multi-DOF Motion Sensing}
\titlenote{Author's accepted manuscript. Accepted for publication in UIST '26.}

\author{Xiang Chang}
\affiliation{%
  \institution{MIT CSAIL}
  \city{Cambridge, MA}
  \country{USA}}
\affiliation{%
  \institution{Tianjin University}
  \city{Tianjin}
  \country{China}}
\email{cassiusxiang@gmail.com}

\author{Haiyang Yan}
\affiliation{%
  \institution{Tianjin University}
  \city{Tianjin}
  \country{China}}
\email{yanhaiyang1231@tju.edu.cn}

\author{Stefanie Mueller}
\affiliation{%
  \institution{MIT CSAIL}
  \city{Cambridge, MA}
  \country{USA}}
\email{stefanie.mueller@mit.edu}

\author{Jiaji Li}
\authornote{Corresponding author.}
\affiliation{%
  \institution{MIT CSAIL}
  \city{Cambridge, MA}
  \country{USA}}
\affiliation{%
  \institution{Zhejiang University}
  \city{Hangzhou}
  \country{China}}
\email{jiaji@mit.edu}

\begin{abstract}
We present X-Hinges, a design and fabrication method for self-sensing compliant mechanisms based on multi-material FDM 3D printing. By co-printing two conductive filaments of different conductivities within a compliant body, we embed resistive sensing elements directly during fabrication without post-assembly, enabling continuous motion sensing across multiple degrees of freedom in a single print. The structure supports three degrees of freedom, each equipped with a dedicated sensing element configuration for \red{multi-DOF motion estimation}. We develop a precision data acquisition system and data-driven regression models that enable continuous, real-time motion sensing. We also introduce an interactive design tool for customizing the geometry, mechanical properties, degrees of freedom, and sensing configurations of X-Hinges. The tool also supports augmenting existing 3D models with self-sensing structures, endowing ordinary objects with continuous multi-DOF sensing capabilities. Finally, we present a set of application examples demonstrating the capability of X-Hinges for fabricating personalized interactive interfaces.

\end{abstract}



\begin{teaserfigure}
\begin{center}
  \includegraphics[width=1\textwidth]{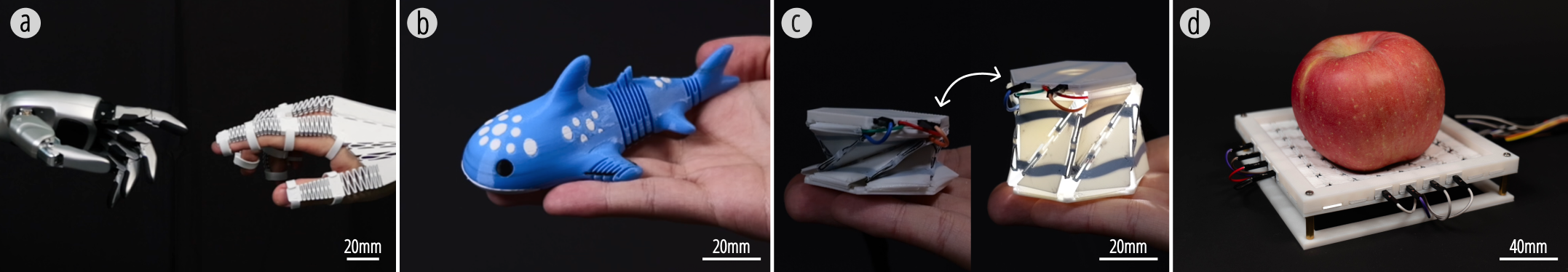}
\end{center}  
    \caption{X-Hinges applications: (a) A moment of connection—self-sensing glove and robotic hand in mirror symmetry. (b) Multi-DOF tangible game controller. (c) Self-sensing Kresling origami lamp. (d) Tactile sensing matrix for object recognition.}
  \vspace*{10pt}
  \Description{Four X-Hinges application examples on a black background. (a) A white, self-sensing 3D-printed glove faces a black-and-silver robotic hand, with their index fingers nearly touching; a 20 mm scale bar is shown. (b) A palm-sized blue orca controller has compliant ribbed joints at its fins and tail; scale bar 20 mm. (c) A Kresling-inspired lamp is shown collapsed and extended on a person's palm, with a curved arrow indicating the transition; scale bar 20 mm. (d) A red apple rests on a square white tactile sensing matrix connected by wires; scale bar 40 mm.}
  \label{fig1}
  \label{fig:teaser}
\end{teaserfigure}

\maketitle
\section{INTRODUCTION}
Self-sensing deformable interfaces are rapidly expanding the possibilities in HCI by turning everyday objects into responsive, self-aware devices. These interfaces have emerged as a prominent research area in HCI, primarily because they enrich physical interfaces with enhanced interactivity and responsive capabilities \cite{10.1145/3322276.3322347, 10.1145/3126594.3126652, 10.1145/2642918.2647405}. Alongside these sensing advancements, rapid progress in digital fabrication techniques has made the construction of complex deformable sensing interfaces far more accessible, supporting rapid iteration and highly personalized designs.

3D printing endows physical structures with rich material properties, allowing sensing functionality to be incorporated into movable objects during fabrication. Early approaches combined 3D-printed structures with manually attached discrete sensors to achieve deformation sensing \cite{10.1145/3472749.3474733, 10.1145/3491102.3501951, 10.1145/3313831.3376136}, but this introduced assembly complexity and limited design flexibility. With advances in multi-material printing and conductive filaments, researchers have moved toward fully integrated fabrication—embedding conductive paths, lattice structures, and metamaterial cells directly into printed objects to sense touch, deformation, and motion without post-assembly \cite{PrintPut,Capricate,Single-Stroke,network,bae2025computational}. These approaches eliminate manual assembly and unify form and sensing within a single print process.

However, despite these advances, two fundamental challenges remain unresolved in 3D-printed sensors: (1) independent \textbf{Multi-axis} sensing and (2) \textbf{Continuous}, high-resolution signal capture. MetaSense \cite{10.1145/3472749.3474806} is limited to detecting in-plane deformation and requires manually inserted \red{bare nickel-chromium wire} as part of its circuit; LattiSense \cite{10.1145/3623263.3623361} successfully senses the magnitude of deformation in printed \red{lattice} structures, but its deformation modes are oriented toward specific single-axis deformations, making simultaneous multi-axis displacement sensing impossible. XSpine \cite{10.1145/3746058.3758350, 10.1145/3772318.3791317} attempts to address this by combining multiple single-axis bending units, yet its contact-based sensing yields only discrete signals, failing to capture the continuous nature of deformation. Enabling continuous sensing across multiple axes simultaneously remains a critical and unsolved challenge.

To address these challenges, we present X-Hinges, a design and fabrication method for 3D-printed compliant structures with continuous, multi-axis self-sensing, all realized in a single multi-material FDM print process. The key insight is that by embedding two conductive materials of contrasting conductivities, namely a high-resistance sensing element ($\rho \approx 1.23\times10^4\ \Omega\cdot\text{cm}$) and a low-resistance conductive trace ($\rho \approx 8.75\ \Omega\cdot\text{cm}$), into a single compliant body, each axis of motion can be assigned a dedicated sensing element. This dual-material architecture directly addresses the scalar ambiguity of single-material approaches, \red{enabling simultaneous sensing across multiple axes with reduced cross-axis coupling}. To further resolve the challenge of continuous sensing, we develop a dedicated data acquisition system and train data analysis models on extensive experimental samples, achieving real-time quantitative perception of deformation. Beyond the sensing system, we investigate how compliant mechanism parameters govern mechanical behavior, and propose a design method for tuning mechanical properties to support personalized interfaces. To streamline the development workflow, we provide an interactive design tool that enables users to configure sensing and motion behaviors according to specific requirements \red{by selectively constraining degrees of freedom within a compliant structure}, and supports embedding self-sensing structures into existing 3D models. Finally, we demonstrate the versatility of X-Hinges through a set of application examples spanning robotics, personalized manufacturing, and intelligent sensing.
In general, our contributions are as follows:
\begin{itemize}
\item A design and fabrication approach that embeds two conductive materials of contrasting conductivities into compliant mechanisms via multi-material FDM printing, \red{enabling continuous sensing of multiple degrees of freedom within a single printed structure.}
\item A dedicated sensing pipeline that couples controlled excitation with low-noise current readout, enabling \red{high-resolution resistance measurement in high-resistance printed sensors and data-driven multi-axis motion estimation.}
\item An interactive design tool that allows users to design the mechanical behavior and sensing layout of X-Hinges, and supports embedding self-sensing structures into existing 3D models to augment objects with deformation awareness.
\item A set of application examples across robotics, personalized fabrication, and intelligent sensing, demonstrating how X-Hinges serve as a general building block for interaction with deformable physical interfaces.
\end{itemize}
\section{RELATED WORK}

\begin{figure*}[h]
    \centering
    \includegraphics[width=0.96\linewidth]{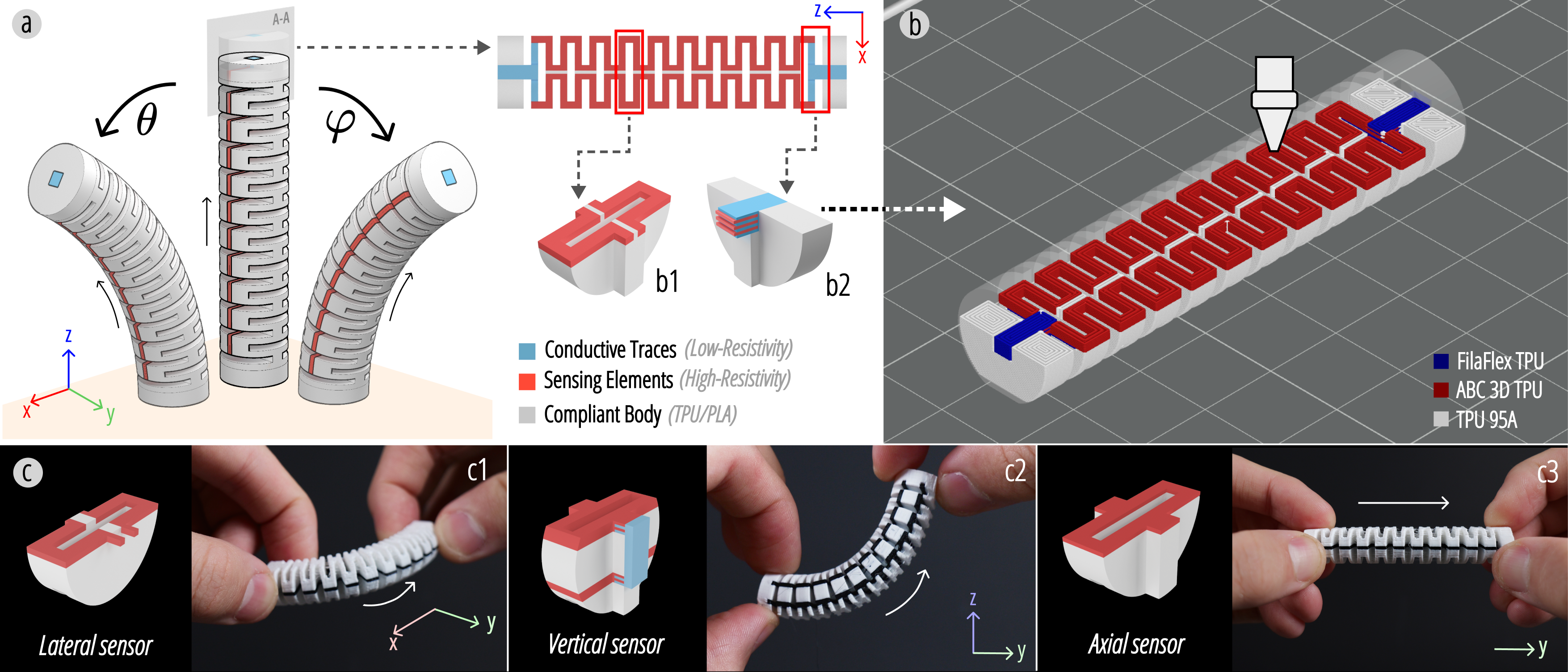}
    \caption{Composition of X-Hinges: (a) The three primary components: compliant body, sensing elements, and conductive traces;(b) The single-step multi-material FDM printing process, realized using three distinct filaments;(c) Three sensing element configurations targeting lateral, vertical, and axial deformation, along with their corresponding circuit designs.}
    \Description{Overview of the X-Hinges structure, fabrication, and sensing layouts. (a) A segmented cylindrical joint bends about two orthogonal angles, theta and phi. A cutaway and enlarged side view identify the gray nonconductive compliant body, red high-resistivity sensing elements, and blue low-resistivity conductive traces; adjacent sensing structures are separated along the z direction. (b) A slicer-style rendering shows all three materials deposited together in one FDM print: white TPU 95A body, red ABC3D conductive TPU sensing elements, and blue Filaflex TPU traces. (c) Schematics and photographs show three layouts: symmetric left-right elements for lateral bending, vertically offset elements for vertical bending, and a centerline element for axial stretching or compression.}
    \label{method1}
\end{figure*}

\subsection{\red{Sensing in Deformable Structures}}
Early physical interfaces in HCI were predominantly composed of rigid, mechanically assembled components—buttons, knobs, sliders, and touchscreens—that constrained users to a limited set of discrete input actions~\cite{10.1145/3322276.3322347}. Tangible User Interfaces advanced this paradigm by physically coupling digital information with graspable physical objects~\cite{10.1145/258549.258715}, yet the underlying structures remained rigid, and sensing was typically achieved through off-the-shelf discrete sensors (potentiometers, force-sensitive resistors, switches) manually mounted within rigid housings.

This has spurred a diverse body of work on sensing deformable interfaces. For example, FlexSense~\cite{10.1145/2642918.2647405} demonstrated how printed piezoelectric sensors can detect complex deformations on flexible surfaces. Similarly, iSoft~\cite{10.1145/3126594.3126654} developed customizable soft sensors using conductive porous silicone materials that support real-time contact and stretching sensing. SmartSleeve~\cite{10.1145/3126594.3126652} proposed a deformable textile sensor that can be seamlessly embedded into garments to enable wearable fabric-based interaction. More recently, Pointner et al.~\cite{10.1145/3746059.3747733} introduced embroidered resonant circuits for textile-based inductive pressure sensing.

Despite the excellent performance of these approaches, most existing works still rely on multi-step fabrication and manual assembly of discrete sensor components, which increases production complexity and limits design flexibility~\cite{IMU}. Retrofitting sensors onto pre-fabricated structures often compromises the intended deformation behavior by introducing undesired stiffness and mass. These constraints make rapid personalized customization of interactive devices particularly challenging.

Building upon the current research landscape, we propose X-Hinges, a self-sensing compliant structure fabricated in a single multi-material FDM print process. \red{Unlike conventional approaches that rely on manually assembled sensing layers and components, X-Hinges embeds sensing capability directly into the structural body in a single print, eliminating the bulk, mass imbalance, and deformation distortion introduced by add-on components.}

\subsection{Compliant Mechanisms for Deformable Interfaces}
The study of compliant mechanisms (CMs) traces back to Burns~\cite{burns1965kinetostatic}, who systematically analyzed flexible-link mechanisms. With advances in computational capabilities and analytical methods~\cite{howell2013compliant}, CMs have since been widely adopted across engineering disciplines. Notable examples include the Lamina Emergent Torsional (LET) joint~\cite{JACOBSEN20092098}, which enables large-angle out-of-plane rotation from a single planar layer, and MEMS-oriented compliant design~\cite{kota2001design} for assembly-free micro-scale precision motion control.

In parallel, a growing body of research has sought to leverage CMs within HCI to deliver customizable physical interaction experiences\cite{x-bridges_2022,all-in-one}. Megaro et al.~\cite{10.1145/3072959.3073636} presented an automated computational tool for converting rigid mechanisms into 3D-printable compliant mechanisms. Li et al.~\cite{Xstrings,Y-zipper} developed integrated cable-driven mechanisms for actuating compliant joints, as well as zipper mechanisms that transform compliant strips into rigid, load-bearing supports. X-Hair~\cite{X-hair} achieves compliant and fine-hair-like structures for aesthetic and functional purposes. FlexHaptics~\cite{10.1145/3491102.3502113} introduced a design methodology for passive haptic input devices based on planar compliant structures. CompAct~\cite{10.1145/3706598.3714307} supports user-defined force--displacement behaviors through multi-joint interconnected compliant networks.

However, existing research has largely focused on mechanical design, leaving sensing potential underexplored. Building upon the LET Joint~\cite{JACOBSEN20092098}, we develop a self-sensing compliant mechanism fabricated in a single FDM print, with dedicated sensing structures for each degree of freedom to enable \red{continuous multi-DOF motion estimation with reduced cross-axis coupling.}

\subsection{Conductive Structures for Deformable Sensing}

Researchers have explored a variety of easy-to-fabricate conductive structures for deformable sensing. Conductive wire-based approaches embed piezoresistive wires into flexible models~\cite{10.1145/2858036.2858354} or perform impedance measurements on a single wire to support multimodal input~\cite{10.1145/3411763.3451547}, but are generally limited to sensing along a single predefined direction. Foam-based structures offer greater geometric freedom: FoamSense~\cite{10.1145/3126594.3126666} senses compression, bending, twisting, and shearing via conductive-ink-impregnated foam; BendID~\cite{10.1145/2632048.2636092} combines conductive foam with fabrics for localized identification; and Aguilar-Segovia et al.~\cite{eduardo_aguilar-segovia_parametric_2025} developed customizable capacitive sensors using FDM-parameterized dielectric structures. However, these methods rely on specialized materials that are difficult to integrate into standard 3D printing workflows.

Researchers have further explored fully printable conductive structures, including capacitive sensing~\cite{10.1145/3025453.3025663}, resistive lattice structures~\cite{10.1145/3623263.3623361}, and contact-based discrete sensing~\cite{10.1145/3746058.3758350}. These approaches achieve fully integrated fabrication without manual assembly or post-processing.

However, these approaches share a fundamental limitation: relying on a single conductive material whose scalar resistive response cannot distinguish deformations along different axes. Prior works therefore either constrain motion to a single direction by design, or sacrifice continuous sensing for discrete detection. \red{Simultaneous continuous estimation of multiple degrees of freedom within a single 3D-printed compliant mechanism remains underexplored.}

To address this, we propose a compliant sensing structure integrating two conductive materials of contrasting conductivities, assigning dedicated sensing configurations to each axis of motion, enabling continuous perception of all three motions within a single printed structure.

\section{X-HINGES MECHANISM}

In this section, we describe the design and sensing principles of X-Hinges. The core of X-Hinges is a three-material architecture, motivated by three practical problems common in 3D-printed sensing structures: (1) when traces and sensing elements share the same material, trace resistance dominates the circuit and weakens the deformation signal; (2) traces deform together with the structure, introducing noise that is difficult to separate from the target signal; (3) conductive material in direct contact with skin introduces body resistance that interferes with measurement. 
\subsection{Structure Composition}
X-Hinges consists of three primary components (Figure~\ref{method1}a): (1) sensing elements printed with high-resistivity conductive filament, maximizing the proportion of signal contributed by deformation in the target region; (2) conductive traces printed with low-resistivity conductive filament, minimizing signal variation caused by unintended trace deformation; (3) a compliant body fabricated from non-conductive filament, isolating the sensing elements from external skin contact. This three-material architecture can be fabricated on any consumer-grade multi-material FDM printer, including the Prusa XL, Bambu H2D, and Snapmaker U1.

\vspace{0.1cm}
\noindent\textbf{Sensing Elements:}
The sensing elements leverage the piezoresistive properties of conductive filaments to enable continuous deformation sensing. We adopt the Lamina Emergent Torsional (LET) compliant joint structure~\cite{JACOBSEN20092098} and fabricate these elements using a high-resistivity conductive TPU filament (ABC3D ESD Grade, $\rho \approx 1.23 \times 10^{4}~\Omega \cdot \text{cm}$). The high resistivity ensures that deformation-induced resistance changes constitute a large proportion of the total circuit resistance, preserving signal strength and sensitivity. By evenly distributing the input force across multiple compliant segments, the structure converts the applied load into uniform torsional and tensile strains, inducing predictable and proportional resistance changes across the full deformation range. Furthermore, by orienting sensing elements along different axes within a single compliant body, each element responds primarily to its target motion direction while remaining insensitive to orthogonal motions, enabling \red{multi-DOF motion estimation} from a \red{single} printed structure, as detailed in Section~\ref{sensing_interface}.

\vspace{0.1cm}
\noindent\textbf{Conductive Traces:}
The conductive traces are printed with a low-resistivity conductive filament (Filaflex, $\rho \approx 8.75~\Omega \cdot \text{cm}$), approximately $10^{3}$ \red{times as conductive as} the sensing elements. This contrast ensures that even when the traces undergo unintended deformation during operation, the resulting resistance change remains negligible relative to that of the sensing elements, effectively isolating the measurement from transmission noise. We further propose an optimized interfacing structure in Section~\ref{evaluation_4} to ensure reliable electrical connection between the two filaments.

\vspace{0.1cm}
\noindent\textbf{Compliant Body:}
The compliant body is fabricated from non-conductive TPU, fully encapsulating the sensing elements to prevent direct skin contact. This ensures that body resistance does not enter the sensing circuit, maintaining measurement accuracy regardless of how the structure is handled. To further enable end-to-end personalization, we present an interactive design tool in Section~\ref{design_tool} that streamlines the configuration of motion primitives and sensing configuration for custom-shaped X-Hinges.

\subsection{ Configuration Layout for Different Motions}\label{sensing}
The three-material architecture establishes the physical foundation for deformation sensing in X-Hinges. However, realizing accurate multi-DOF sensing introduces a concrete technical challenge: cross-sensitivity—when multiple degrees of freedom are actuated simultaneously, each sensing element may respond to deformations from multiple axes, making it difficult to isolate individual motion components.

To address this, X-Hinges draws on the differential measurement principle of the Wheatstone bridge~\cite{hoffmann1974applying}, implementing the bridge logic directly through the 3D-printed physical geometry. We designed dedicated sensing layouts for each of the three motion axes (Figure~\ref{method1}c), maximizing signal response along the target axis while suppressing interference from orthogonal directions at the structural level. Cross-axis interference is reduced to as low as 8.2\% through geometric design and differential signal processing.

When X-Hinges deforms, its structure converts load into uniform strains in the sensing elements. The piezo-resistive behavior of conductive TPU directly modulates electrical resistance in response to mechanical strain. To capture this change, we apply a constant excitation voltage (\(5\,\mathrm{V}\)) and monitor the resulting feedback current $I_f$ through a transimpedance amplifier (Section~\ref{sensing-hardware}). Because the sensing elements have significantly higher resistance than the conductive traces, the traces' own resistance and any noise from their unintended deformation remain negligible, ensuring that the measured signal accurately reflects the joint's deformation state.

\vspace{0.1cm}
\noindent\textbf{Lateral Sensing Configuration: }
Two sensing elements are placed symmetrically on the left and right sides of the joint's centerline. Subtracting their signals isolates lateral bending while rejecting vertical and axial noise through common-mode cancellation.

\vspace{0.1cm}
\noindent\textbf{Vertical Sensing Configuration: }
Two sensing elements are positioned at different heights along the joint's vertical axis. Their differential output captures vertical bending while filtering out lateral and axial components, ensuring orthogonal decoupling.

\vspace{0.1cm}
\noindent\textbf{Axial Sensing Configuration: }
A single sensing element is placed along the structural centerline, vertically centered between the top and bottom edges of the joint. At this symmetric position, lateral and vertical bending induce equal and opposite strains that cancel in the net resistance change, leaving only axial compression or elongation as the dominant signal.

\vspace{0.1cm}
By selectively embedding one or more sensing configurations within a single printed structure, X-Hinges can be tailored to monitor only the required degrees of freedom. \red{When multiple configurations are integrated, a minimum spacing of $z=0.6\,\mathrm{mm}$ is maintained to prevent interference between adjacent sensing structures.} \red{Combining all three enables simultaneous sensing of three motion components with reduced cross-axis coupling, supporting continuous multi-axis motion estimation in a single printed structure.}

\begin{figure*}[h]
    \centering
    \includegraphics[width=0.96\linewidth]{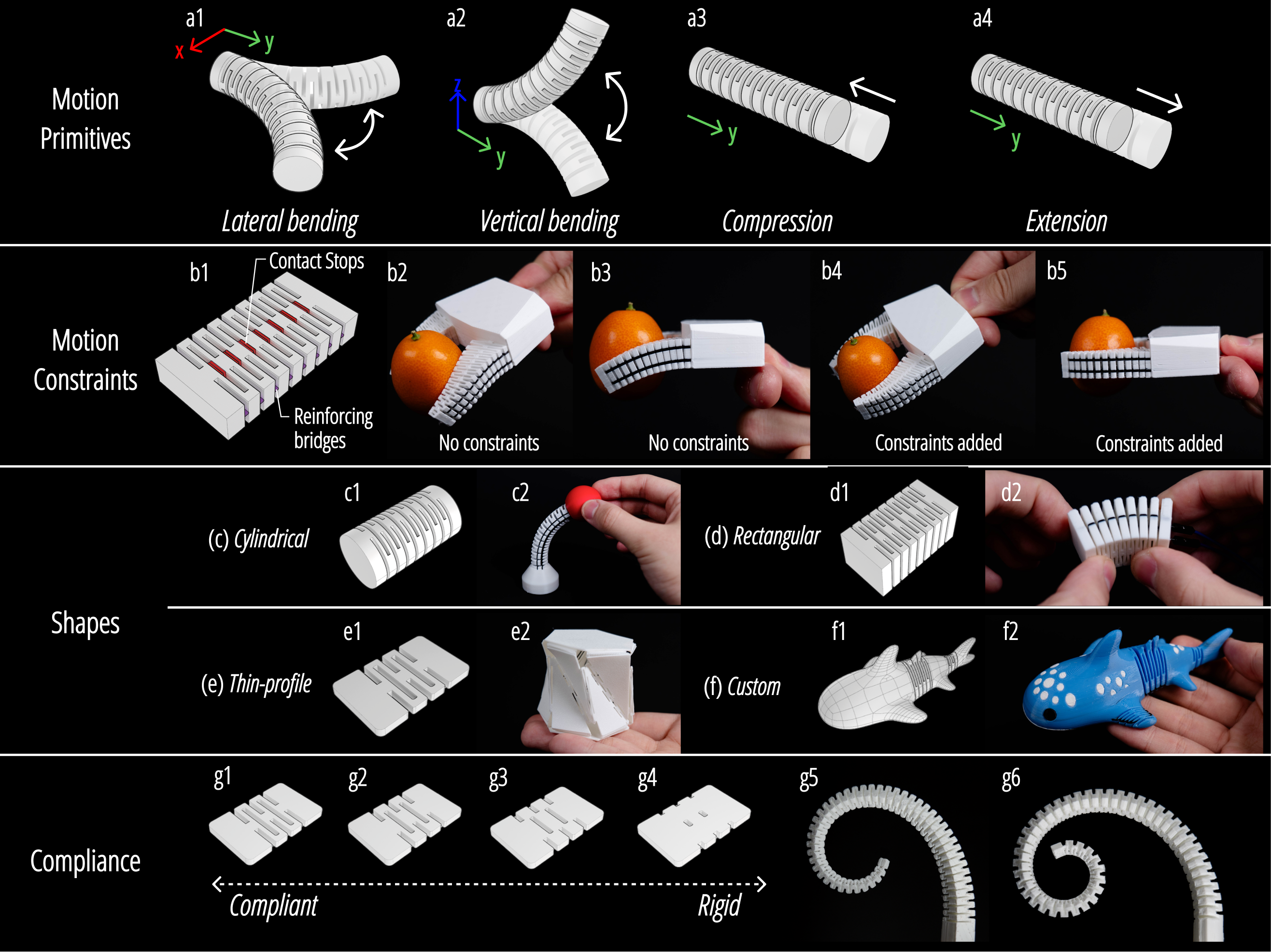}
    \caption{Design space of X-Hinges, spanning motion primitives, motion constraints, geometric shapes, and mechanical compliance, shown with 3D models and photographs of representative structures and behaviors.}
    \Description{Design space organized into four rows. The motion-primitives row shows lateral bending, vertical bending, compression, and extension. The motion-constraints row shows contact stops and reinforcing bridges, followed by a gripper without constraints bending undesirably around an orange and a constrained gripper maintaining its intended orientation. The shapes row pairs models and printed examples of cylindrical, rectangular, thin-profile, and custom geometries, including a folded form and an orca. The compliance row progresses from narrow, flexible bridges to wide, rigid bridges; two tendon-driven specimens demonstrate that a more compliant structure forms a tighter spiral while a stiffer structure forms a looser coil.}
    \label{Design_Space}
\end{figure*}

\section{DESIGN SPACE}\label{physical_interface}

A fundamental challenge in the design of 3D-printed compliant mechanisms is achieving controllable motion and tunable mechanical behavior within a single body. To address this, we formalize the X-Hinges design space through three integrated dimensions: \textbf{Motion Primitives}, \textbf{Geometric Form}, and \textbf{Mechanical Stiffness}. These dimensions allow users to independently tailor the kinematic, geometric, and mechanical properties of an X-Hinges structure to meet specific application requirements.

\subsection{Subtractive Motion Primitives}
X-Hinges adopts a subtractive kinematic approach to motion design. Unlike traditional additive mechanisms that stack discrete single-DOF joints, we define three innate motion primitives \textit{lateral bending}, \textit{vertical bending}, and \textit{axial translation}—as the fundamental degrees of freedom inherent to the structure. These three axes form a holistic kinematic basis, providing a tri-axial workspace within a single printed body.

Customizing an X-Hinges instance involves selectively constraining these innate primitives. By introducing targeted structural modifications, designers can "prune" redundant degrees of freedom to achieve precise, task-specific movement:

\vspace{0.1cm}
\noindent\textbf{Rotational Constraints:} To suppress one bending primitive while preserving the other, reinforcing bridges are introduced perpendicular to the target axis (Figure~\ref{Design_Space}b). This increases stiffness in the constrained direction by up to 22.5$\times$, converting a multi-axis joint into a stabilized, single-DOF vertical or lateral hinge.

\vspace{0.1cm}
\noindent\textbf{Translational Constraints:} To decouple axial motion from rotational bending, contact stops are embedded within the compliant segments. \red{These stops engage under axial compression to limit compression while remaining disengaged during bending. Axial extension is limited by a centerline bridge that carries tensile load like a string during stretching, with minimal effect on bending.}

\vspace{0.1cm}
By treating motion as a subtractive process, X-Hinges can be tailored to match exact kinematic requirements—such as a stabilized gripper (Figure~\ref{Design_Space}b2-5) that resists lateral wobbling or a compression-proof folding joint—without increasing the physical complexity of the assembly.

\subsection{Geometric Form and Integration}
The cross-sectional morphology of X-Hinges is highly adaptable, enabling seamless integration into various physical form factors. We provide three standardized geometric primitives—\textit{thin-profile}, \textit{rectangular}, and \textit{cylindrical} (Figure~\ref{Design_Space}c-e)—each optimized for specific mechanical and spatial constraints. 

\vspace{0.1cm}
\noindent\textbf{Rectangular:} profiles provide flat mounting surfaces and enhanced structural rigidity, suitable for building blocks or modular mechanical assemblies where precise alignment is required.

\vspace{0.1cm}
\noindent\textbf{Cylindrical:} forms are designed to match the organic shape of human joints or robotic limbs, facilitating the creation of ergonomic wearables and biomimetic structures.

\vspace{0.1cm}
\noindent\textbf{Thin-profile:} geometries minimize the physical footprint, making them ideal for origami-inspired surfaces or flexible strips that require high bendability with minimal bulk.

\vspace{0.1cm}
Beyond these standard forms, X-Hinges can be customized to follow arbitrary 3D contours. This flexibility allows the mechanism to be embedded directly into complex host geometries via our interactive design tool (\red{Figure~\ref{design_tool_generate}}), transforming passive 3D models into self-sensing interactive devices.

\subsection{Mechanical Stiffness}
The mechanical stiffness of X-Hinges governs its passive tactile feedback and load-bearing capacity, tunable from highly flexible to nearly rigid. This allows users to match the mechanical feel of an interface to its intended interaction—from soft, \red{low-stiffness} joints for wearable finger-tracking to stiff, spring-loaded segments for load-bearing applications.

This tunability is primarily achieved by modulating the internal geometry of the compliant bridges. As characterized in our evaluation (Appendix~\ref{evaluation_0}), varying the bridge width $w$ alone allows the bending stiffness to span more than two orders of magnitude (up to a 470$\times$ increase).

To demonstrate this localized stiffness control, we fabricated tendon-driven tentacle-like structures with varying bridge width distributions (Figure~\ref{Design_Space}g5-6). Under the same actuation force, segments with narrower bridges yield a tight spiral curvature, while those with wider bridges maintain a loose coil, illustrating how spatially graded compliance can be used to program complex, non-uniform deformation behaviors within a single printed body.

\section{SENSING SYSTEM}\label{sensing_interface}
In this section, we introduce the X-Hinges sensing system. The system is designed to address three fundamental challenges commonly encountered in 3D-printed resistive sensors: (1) high-resistance sensing structures produce weak signals that are difficult to measure accurately with traditional readout circuits, limiting continuous real-time sensing; (2) decoding multi-DOF motion from resistive signals is non-trivial due to cross-sensitivity and hysteresis effects; and (3) variations in printing parameters and filament batches lead to unpredictable resistance baseline shifts across different prints.

\subsection{X-Hinges Sensing Hardware} \label{sensing-hardware}
\begin{figure}[h]
  \includegraphics[width=0.96\linewidth]{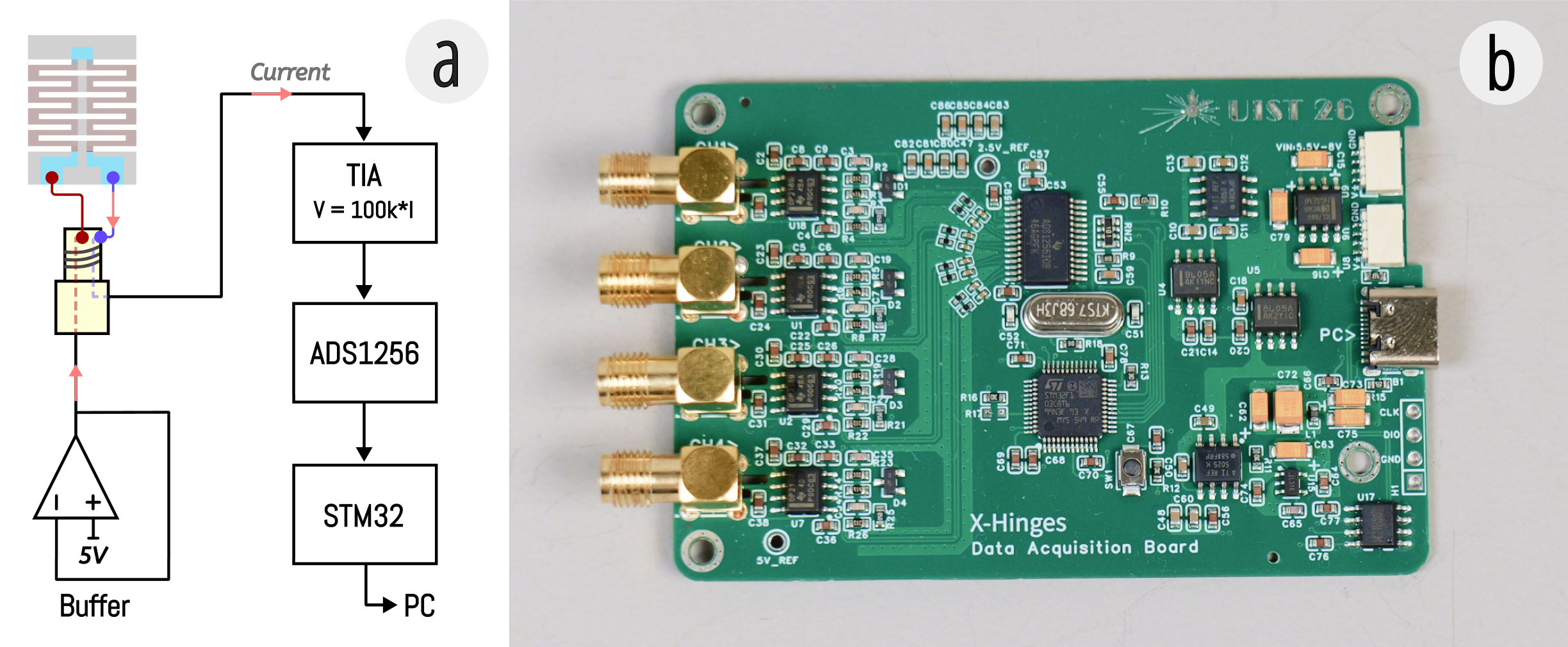}
  \caption{X-Hinges sensing hardware. (a) Block diagram of the sensing circuit, showing the main composition. (b) Photograph of the fabricated sensing hardware PCB.}
  \Description{X-Hinges sensing hardware. (a) A block diagram traces the sensing path: a 5 V buffer excites the printed sensor, feedback current passes through a transimpedance amplifier with output V equals 0.1 megaohm times I, an ADS1256 digitizes the voltage, and an STM32 sends data to a PC. (b) Photograph of the fabricated green X-Hinges Data Acquisition Board, with four gold sensor connectors along the left edge, signal-conditioning components and microcontroller in the center, and power and USB connections on the right.}
  \label{hardware}
\end{figure}

\red{To address weak signals from high-resistance sensing structures, we developed dedicated hardware for real-time resistance acquisition (Figure~\ref{hardware}).}

As shown in Figure~\ref{hardware}a, the hardware pipeline consists of three stages: a voltage follower provides stable excitation, a transimpedance amplifier (TIA) converts the feedback current into voltage, and an ADC (ADS1256) digitizes the signal. The sensing element resistance is then calculated as:
\begin{equation}
R_{sense} = \frac{V_{\text{exc}} \cdot R_{\text{TIA}}}{V_{\text{ADC}}}
\label{eq:resistance}
\end{equation}
where $V_{\text{exc}} = 5V$ is the excitation voltage, $R_{\text{TIA}}$ is the transimpedance gain, and $V_{\text{ADC}}$ is the digitized voltage output.

The hardware achieves a resistance measurement accuracy of 0.01\% full scale with noise below 0.015\% across a 0.1-100\,M$\Omega$ range. We adopt a 50\,Hz sampling rate per channel. A single board supports 4 channels, scalable to 32. Compared with the ESP32 ADC approach, our design improves signal-to-noise ratio by up to 27,415$\times$ for high-resistance sensors. Detailed performance comparison and circuit schematics are provided in Appendices~\ref{hardware_comparison} and~\ref{hardware_sch}, respectively.

\subsection{Multi-DOF Motion Decoding}

\begin{figure}[h]
    \centering
    \includegraphics[width=0.96\linewidth]{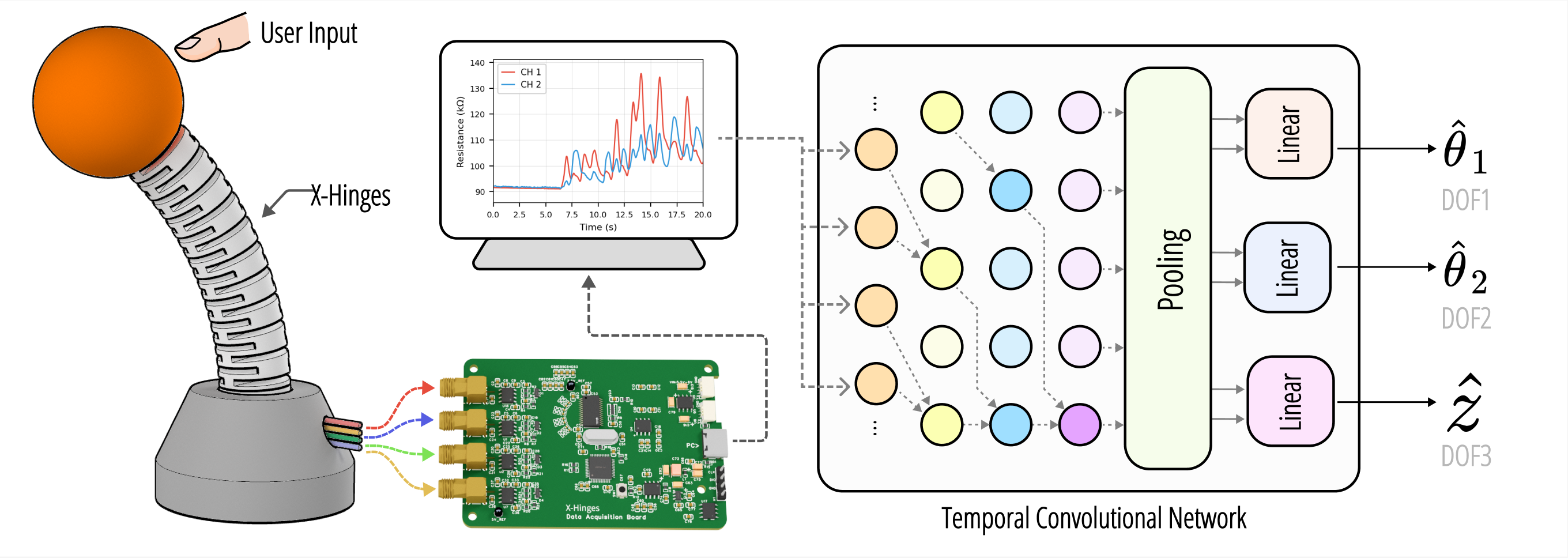}
    \caption{Overview of the multi-DOF deformation decoding pipeline.}
    \Description{Multi-DOF decoding pipeline from physical input to estimated motion. A finger pushes the orange end of a flexible X-Hinges column. Four colored sensor leads connect it to the acquisition board, which sends multi-channel resistance histories to a computer; an example plot shows two channels changing over 20 seconds. The histories enter a Temporal Convolutional Network with stacked temporal feature layers, pooling, and three linear output heads. The outputs are estimates of two bending angles, theta 1 and theta 2, and axial displacement z, corresponding to three degrees of freedom.}
    \label{pipeline}
\end{figure}

\red{Although the sensing configurations in Section~\ref{sensing} reduce cross-axis interference structurally, residual coupling and piezoresistive hysteresis make instantaneous resistance readings
ambiguous. We therefore use a Temporal Convolutional Network (TCN)~\cite{bai2018empirical} to jointly decode multiple motion components from multi-channel resistance histories.}

\red{The decoder contains four residual temporal blocks with dilation rates of 1, 2, 4, and 8. Each block applies two causal one-dimensional convolutions with weight normalization, ReLU activation, dropout, and
a residual connection. Adaptive average pooling aggregates the temporal features, and a linear head outputs the three motion components. Data are divided chronologically into 70\% training, 15\% validation, and 15\% test segments before window construction. The model is optimized using smooth L1 loss and AdamW, with model selection based on validation loss. Figure~\ref{Evaluation_8_2_2}c shows the decoded three-DOF trajectories, with quantitative results reported in Section~\ref{evaluation_integrated} and additional details in Appendix~\ref{dof_decoding_demonstration}.}

\subsection{Interactive Self-Calibration} \label{self-calibration}
While the decoding pipeline above generalizes well within a single instance, signal-to-output mappings are rarely transferable across prints due to variations in print parameters, material batches, and geometric tolerances. Here we introduce an interactive self-calibration mechanism that adapts a pre-trained base model to each newly printed instance through lightweight fine-tuning.

\begin{figure}[h]
    \includegraphics[width=0.96\linewidth]{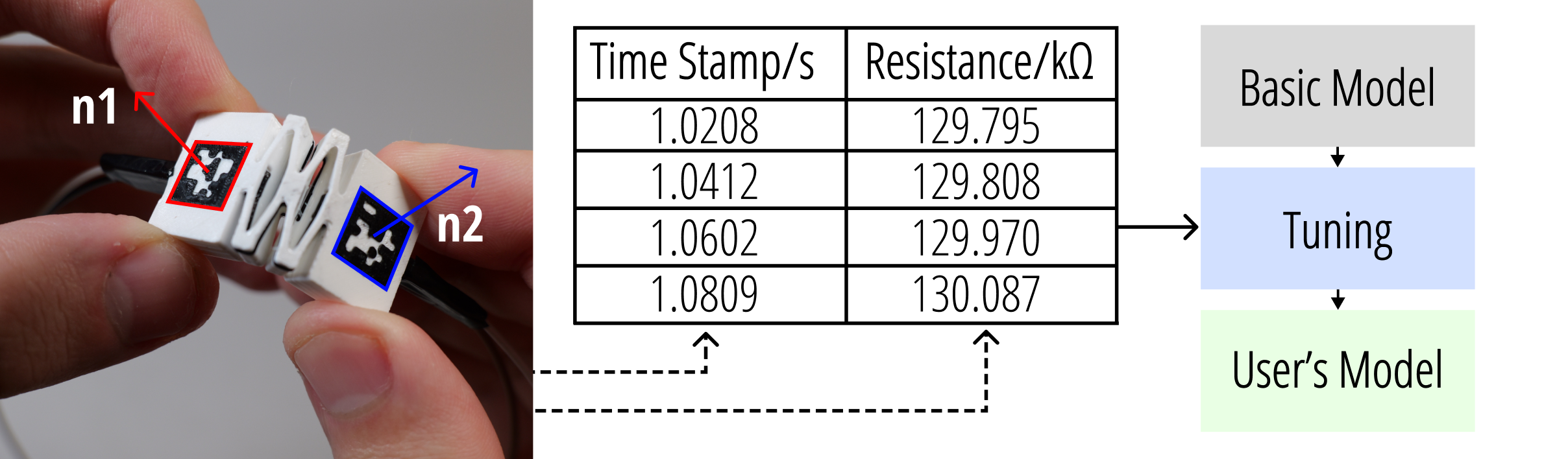}
    \caption{Self-calibration pipeline using co-printed AprilTag markers.}
    \Description{Interactive self-calibration workflow. At left, two co-printed black-and-white AprilTag markers on a bent X-Hinge are outlined in red and blue; arrows label their normal vectors n1 and n2. In the center, camera-derived pose samples are synchronized with resistance samples, illustrated by a two-column time-stamp and resistance table. At right, the labeled pairs feed a tuning stage initialized from a basic model, producing a specimen-specific user's model.}
    \label{calibration_pipeline}
\end{figure}

During calibration, the user freely manipulates the structure while a camera (e.g., smartphone or laptop) tracks the motion. The system automatically pairs resistance readings with camera-derived ground-truth labels. \red{For a new 1-DOF geometry, collecting sufficient data to train the base model takes approximately 8 minutes; for subsequent prints of the same geometry, only a 2-minute calibration run is required for specimen-specific fine-tuning.}

This calibration data is used to: (1) re-fit input normalization parameters to compensate for resistance baseline shifts, and (2) retrain the model's upper layers to adapt to instance-specific characteristics. \red{AprilTag markers can be permanently printed in TPU or temporarily printed in peelable black PLA.}

As shown in Figure~\ref{calibration_pipeline}, two AprilTag markers are co-printed onto the X-Hinges structure. The camera estimates marker poses in real time, and the bending angle is computed as:
\red{
\begin{equation}
    \theta =
    \operatorname{atan2}
    \left(
      \mathbf{a}\cdot
      \left(\mathbf{n}_1\times\mathbf{n}_2\right),
      \mathbf{n}_1\cdot\mathbf{n}_2
    \right),
    \label{eq:calib}
\end{equation}
}
\red{where $\mathbf{a}$ is the unit vector defining the positive bending axis. All vectors are expressed in the same camera coordinate frame. Each camera-derived angle is then time-aligned with the nearest resistance sample to produce labeled training pairs.}

\section{DESIGN TOOL}\label{design_tool}
To streamline the workflow from concept to printable model, we implement an interactive design tool built on Grasshopper and HumanUI within the Rhino environment. Unlike manual CAD workflows that require separately modeling structural geometry, sensing elements, and conductive traces, our tool automates the entire pipeline from geometry input to print-ready file across three stages: \textit{Design CM Structure}, \textit{Modify CM Structure}, and \textit{Generate Sensing Configuration}.

\begin{figure}[h]
    \includegraphics[width=0.96\linewidth]{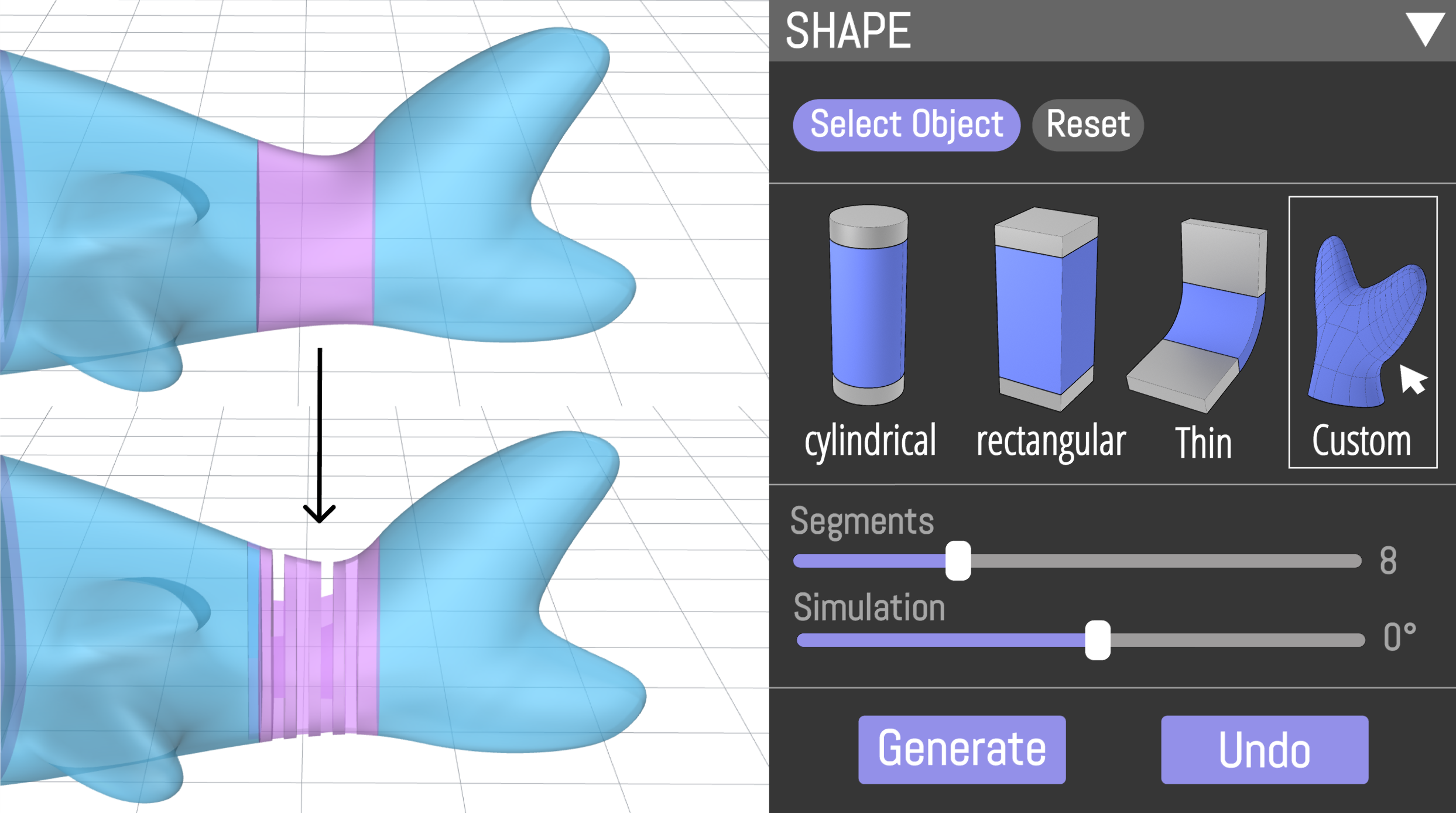}
    \caption{Interactive design tool interface for generating X-Hinges: shape selection, segment configuration, and deformation preview.}
    \Description{Design-tool interface for generating an X-Hinges joint inside a custom object. The left side shows a translucent blue orca model before and after a purple section near the tail is converted into eight compliant segments. The right control panel offers cylindrical, rectangular, thin, and custom shape modes, with Custom selected; sliders set Segments to 8 and Simulation to 0 degrees, and Generate and Undo buttons apply or revert the result.}
    \label{design_tool_generate}
\end{figure}

\begin{figure}[h]
    \includegraphics[width=0.96\linewidth]{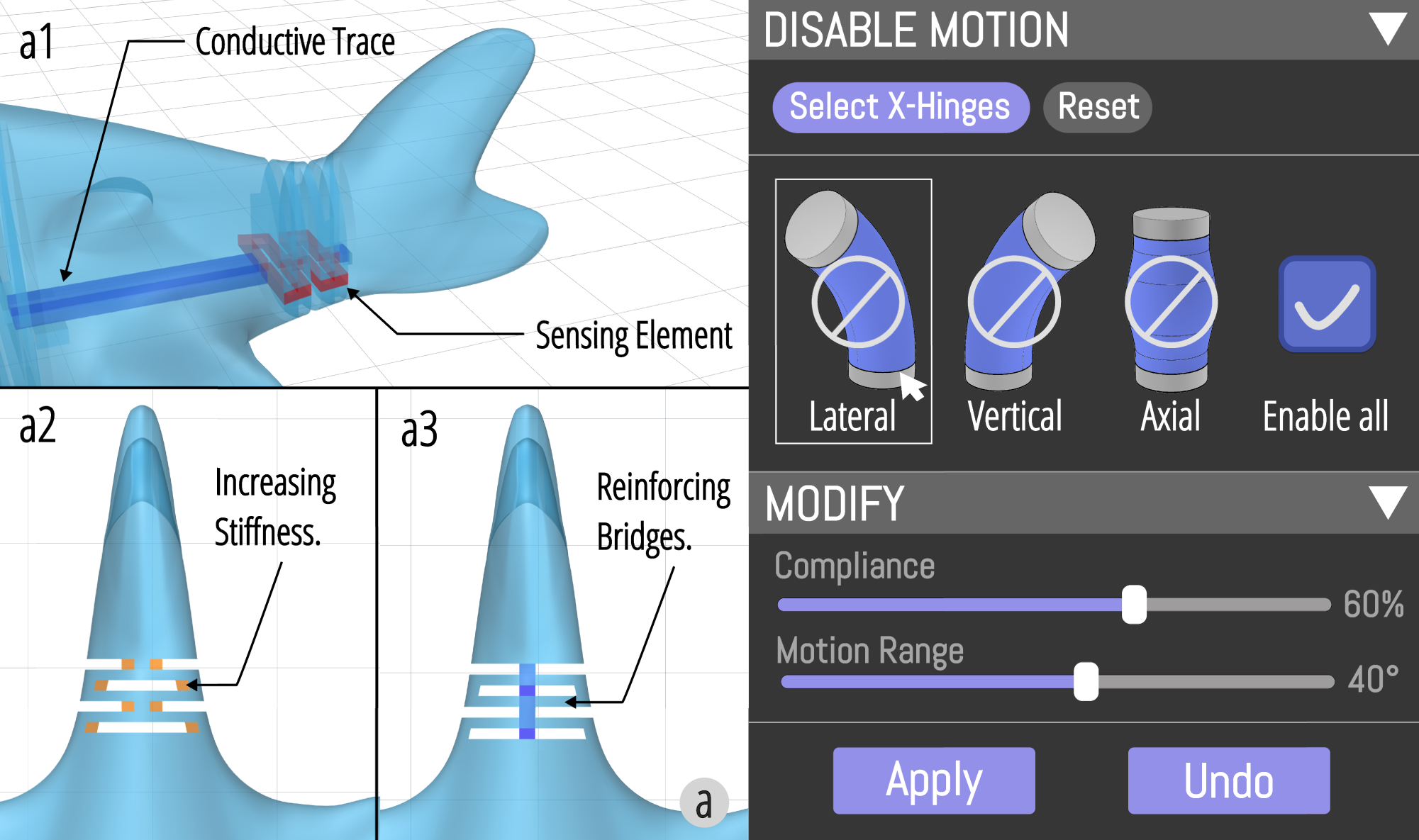}
    \caption{Modification interface for customizing X-Hinges behavior: Modify panel: motion constraints, compliance tuning, and motion range adjustment.}
    \Description{Design-tool interface for modifying motion and sensing behavior in a custom orca model. The upper-left viewport shows a translucent orca with a blue conductive trace running through the body and red sensing elements embedded in the tail joint. Lower-left examples show orange local bridge changes that increase stiffness and a blue center reinforcing bridge. The right panel can disable lateral, vertical, or axial motion or enable all; lateral is selected in the example. Sliders set compliance to 60 percent and motion range to 40 degrees, with Apply and Undo controls.}
    \label{design_tool_modifed}
\end{figure}

\vspace{0.1cm}
\noindent\textbf{Design CM Structure.}
Users can design the compliant mechanism base structure in four shape modes: \textit{cylindrical}, \textit{rectangular}, \textit{thin-profile}, and \textit{custom}. The tool first constructs an interleaved array of rectangular laminae, with segment count controlling the number of compliant units. It then performs a Boolean difference operation to subtract these laminae from the input geometry, leaving behind the compliant body as a movable \red{single-piece compliant structure}. In \textit{custom} mode, users select any existing 3D object in the viewport and specify the target location; the tool automatically embeds the CM base structure at the designated position, endowing ordinary objects with compliant joints in a single step. Figure~\ref{design_tool_generate} shows an orca model transformed by designing compliant joints at its tail and fin positions.

\vspace{0.1cm}
\noindent\textbf{Modify CM Structure.}
Once generated, users can further customize the structure through three controls. The \textit{Compliance} slider maps to the lateral bridge thickness, tuning bending stiffness across a wide range. The \textit{Motion Range} slider controls the axial bridge thickness, adjusting the structure's translational travel. A \textit{Simulation} bar provides real-time deformation preview at any target angle, allowing users to verify mechanical behavior before committing to fabrication. Motion directions can be selectively disabled—lateral, vertical, or axial—with the tool automatically applying the corresponding structural constraints.

\vspace{0.1cm}
\noindent\textbf{Generate Sensing Configuration.}
Upon finalizing the motion design, the tool automatically generates the appropriate sensing element layouts based on the enabled degrees of freedom. For example, if lateral and axial motions are disabled (Figure~\ref{design_tool_modifed}), only the vertical sensing configuration is embedded. When all three axes are active, all three configurations are co-generated within the structure, though this requires a sufficiently large host geometry to accommodate the sensing elements. 

The final model is exported as a \texttt{.3mf} file with pre-assigned material regions, ready for direct import into slicing software. Users can follow the print parameters provided in Appendix~\ref{print} to ensure reliable single-pass multi-material fabrication.

\section{APPLICATIONS}

Building on the sensing method and physical interface design presented above, we demonstrate four application examples that explore the breadth of X-Hinges across different interaction contexts.

\subsection{Self-Sensing Data Glove}

The rapid advancement of embodied AI and robot learning relies on capturing human demonstrations through dexterous teleoperation. However, commercial teleoperation gloves are expensive, cumbersome, and difficult to adapt to diverse hand anatomies. To demonstrate the potential of X-Hinges in personalized human-robot interaction, we designed a wearable teleoperation glove entirely 3D-printed in a single pass.

Our glove provides continuous kinematic tracking across all five fingers. Single-DOF X-Hinges distributed across the metacarpophalangeal (MCP) and interphalangeal (IP) joints capture finger flexion. A multi-DOF X-Hinge at the thumb base captures lateral swing.
The system consists of a dexterous hand (LinkerHand O6) on a robotic arm (OpenArmX). A PICO 4 Ultra tracks the user's wrist pose, which is mapped to the robotic arm through inverse kinematics. The glove acquires sensor signals at \(50\,\mathrm{Hz}\) and decodes them into finger angles using the pipeline from Section~\ref{sensing_interface}, then maps them to robotic hand joints.

The system successfully completed thumb-to-finger opposition, handshake gestures, and two-finger pinch-and-pick tasks, with end-to-end latency of 180-200 ms. This demonstrates how X-Hinges enable rapid prototyping of body-conformal, personalized sensing interfaces as low-cost alternatives to commercial teleoperation gloves, without complex assembly.

\begin{figure}[h]
    \includegraphics[width=0.96\linewidth]{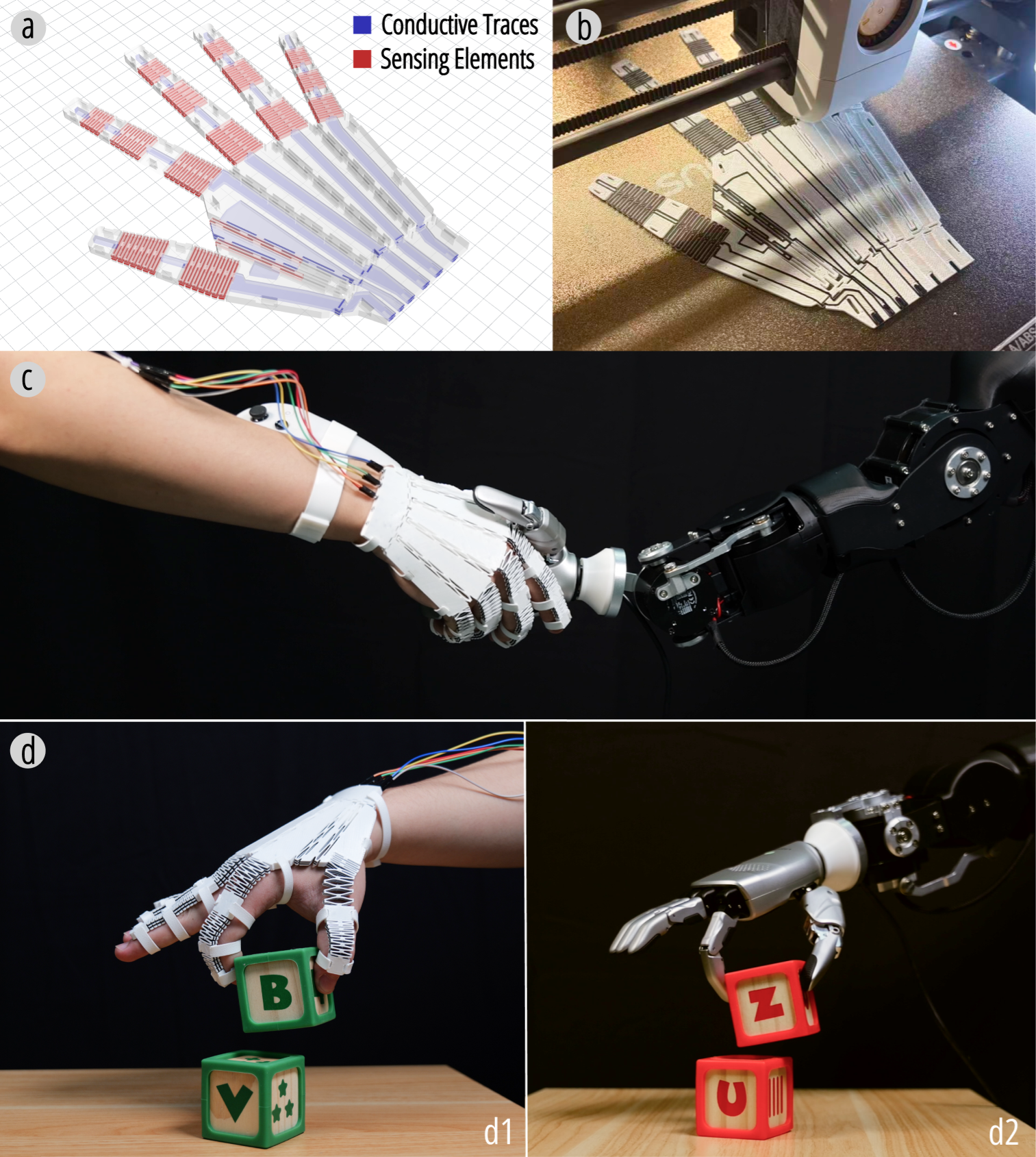}
    \caption{X-Hinge data glove for teleoperation. (a) Circuit layout. (b) \red{Single-piece 3D-printed fabrication}. (c) Human-robot handshake via real-time kinematic transfer. (d) Synchronized grasping via motion mapping.}
    \Description{Self-sensing data glove for robot teleoperation. (a) A CAD view of the glove highlights red sensing elements across the finger joints and blue conductive traces running to the wrist. (b) The flat, single-piece glove is being fabricated on a multi-material FDM printer, with dark conductive paths embedded in the white structure. (c) A wearer using the wired white glove shakes hands with a robotic hand. (d) Two paired images demonstrate synchronized grasping: the wearer pinches and lifts a green letter block, and the robot mirrors the grasp with a red letter block.}
    \label{Application1}
\end{figure}

\begin{figure}[h]
    \includegraphics[width=0.96\linewidth]{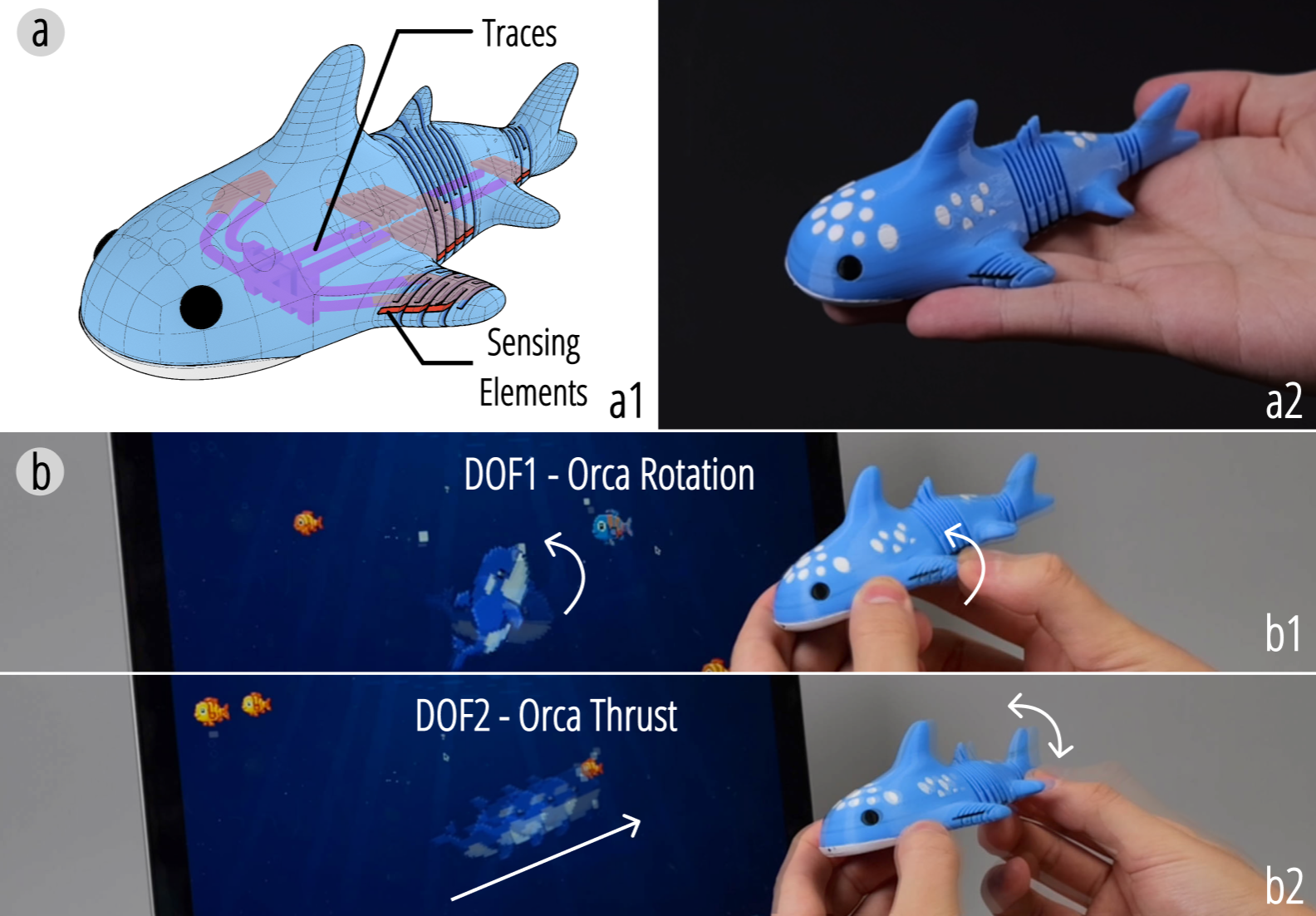}
    \caption{Orca-shaped 2-DOF tangible game controller. (a) Continuous sensing in pectoral fins and tail. (b) Game demo: (b1) fin flexion for yaw; (b2) tail swing for propulsion.}
    \Description{Orca-shaped two-DOF tangible game controller. (a1) A transparent CAD rendering exposes conductive traces through the body and sensing elements at both pectoral fins and the tail base. (a2) Photograph of the palm-sized blue printed controller. (b1) Flexing a pectoral fin rotates the on-screen orca, mapping to the first degree of freedom. (b2) Swinging the tail side to side propels the on-screen orca forward, mapping to the second degree of freedom.}
    \label{Application2}
\end{figure}

\subsection{Tangible Game Controller}

Leveraging the capability of X-Hinges to embed continuous multi-DOF sensing into arbitrary forms, users can design personalized interactive devices that map physical manipulation directly to digital behavior. We demonstrate this with a custom orca\red{-}shaped game controller for a 2D arcade game.

The controller features a 2-DOF sensing architecture mirroring the orca's natural movements. Two X-Hinges elements embedded in the left and right pectoral fins are wired in series; flexing the fins generates differential resistance changes controlling the avatar's yaw. A second sensing axis at the tail base responds to side-to-side swings to modulate forward propulsion.

Unlike conventional controllers requiring memorized button mappings, this design establishes one-to-one correspondence between physical model and
digital counterpart, enabling intuitive operation without instruction. This demonstrates how X-Hinges enable rapid prototyping of expressive tangible
interfaces where physical form becomes the interaction language.

\subsection{Self-Sensing Origami}

Kresling origami structures are valued for their deployability and compact form, yet traditional fabrication methods struggle to embed sensing capability without compromising their folding kinematics. The single-step printing of X-Hinges makes it a natural fit for such structures—the entire Kresling unit is printed flat in a single pass, with a wall thickness of only 1.2~mm and a structural weight of just 13.85~g; the complete system including lamp, battery, and circuit weighs 28.9~g.

\begin{figure}[h]
    \includegraphics[width=0.96\linewidth]{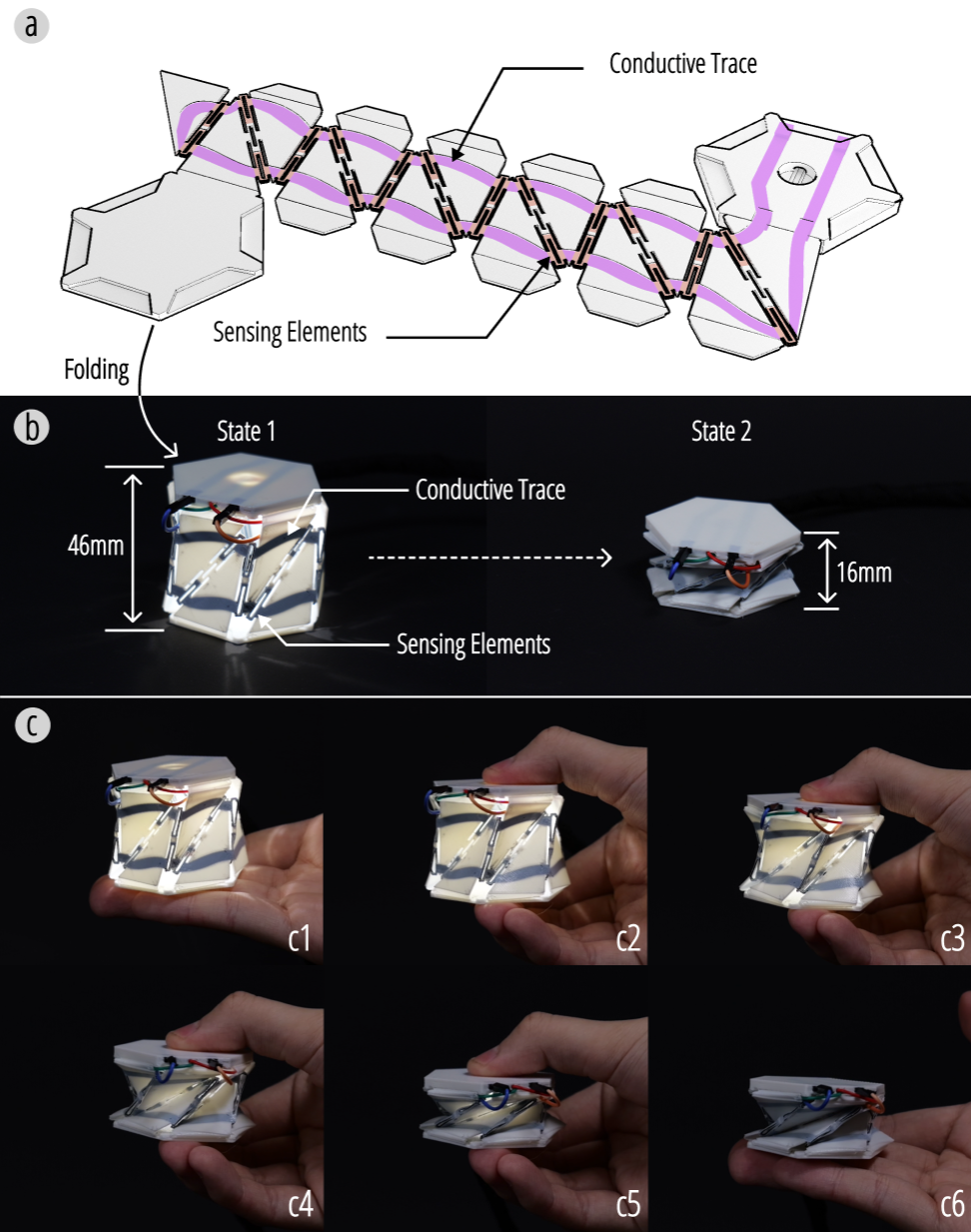}
    \caption{Kresling-inspired lamp. (a) Circuit layout of the \red{unfolded structure printed in a single pass}. (b) \red{Folding} states: extended (46 mm, Bright) and collapsed (16 mm, Off). (c) Continuous dimming during structural collapse.}
    \Description{Kresling-inspired self-sensing lamp. (a) The unfolded, single-pass printed pattern consists of connected polygonal panels with black sensing elements along fold lines and a purple conductive trace linking them. (b) The assembled lamp changes from a 46 mm tall extended, illuminated state to a 16 mm tall collapsed, unlit state. (c1-c6) Six photographs show a hand progressively pressing the lamp from fully extended to fully collapsed; its light continuously dims through the sequence.}
    \label{Application3}
\end{figure}

By tuning the compliance parameters of the embedded X-Hinges, the structure can rest naturally at any folding state without requiring sustained force. X-Hinges embedded along the valley creases continuously capture resistance changes during folding, mapped directly to LED brightness: folding depth controls luminance in a one-to-one correspondence, enabling fluid dimming from full intensity to off.

This application demonstrates how X-Hinges can transform ordinary deployable structures into intelligent interactive interfaces with continuous sensing capability.

\subsection{Tactile Sensing Matrix}

One vision of ubiquitous computing~\cite{lyytinen2002ubiquitous} is to endow everyday surfaces with the ability to recognize objects placed upon them, a capability that is equally relevant to robotic electronic skin and intelligent workbenches. However, existing solutions rely on complex sensor arrays that are costly to customize and deploy. The single-step printing of X-Hinges offers a low-cost path toward this vision: a sensing matrix can be printed in one pass and attached to any surface without additional assembly, endowing robot limbs or everyday objects with tactile perception.

\begin{figure}[h]
    \includegraphics[width=0.96\linewidth]{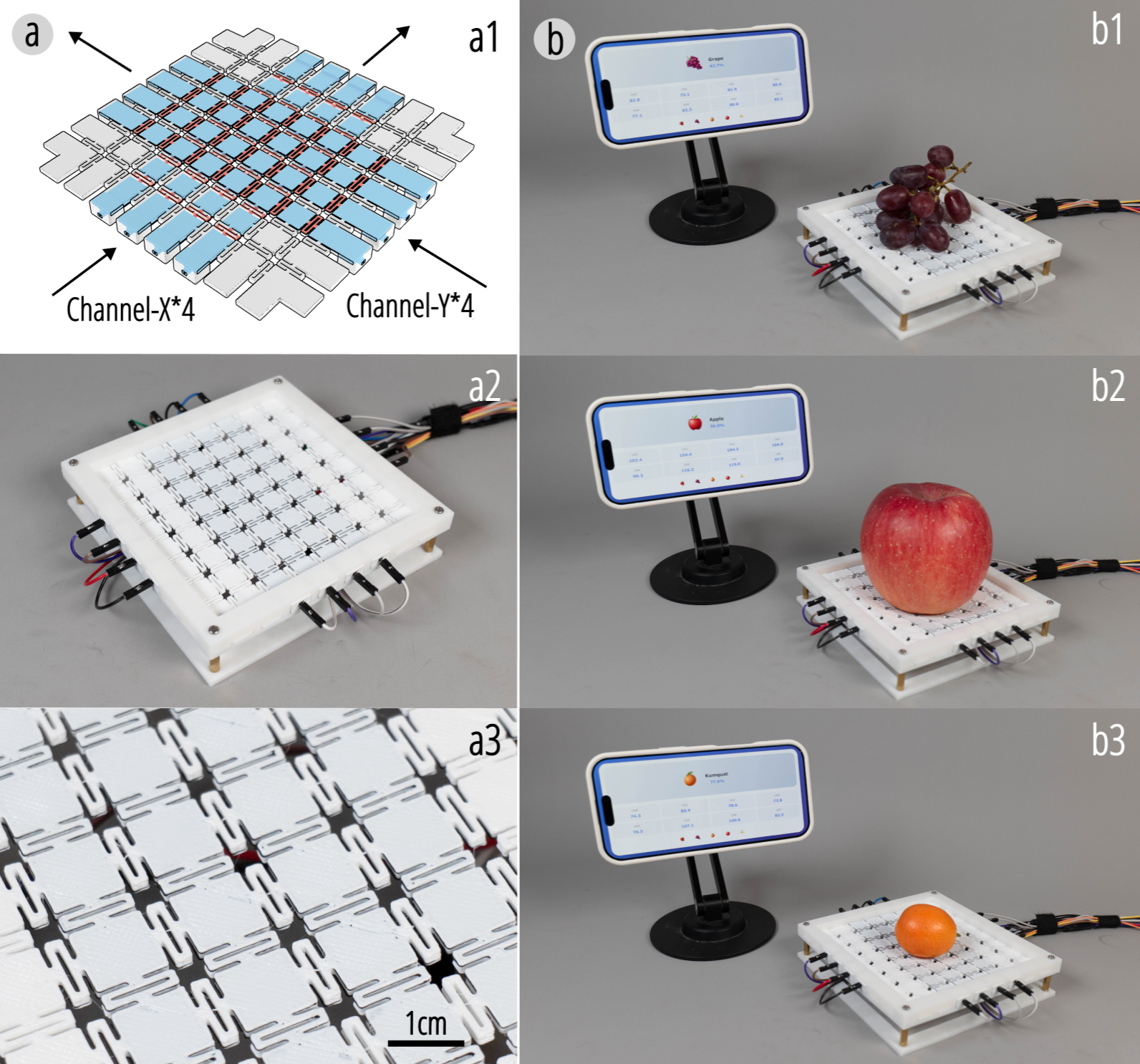}
    \caption{Tactile sensing matrix for object recognition. (a) System architecture: (a1) \red{integrated printed circuit layout}, (a2) experimental setup (b) Classification across diverse objects: Grape (92.7\%), Apple (96.0\%), and Kumquat (77.6\%).}
    \Description{Tactile sensing matrix and fruit-recognition demonstration. (a1) A circuit-layout rendering shows a square grid of compliant cells crossed by four X channels and four Y channels. (a2) Photograph of the wired white matrix in a square frame. (a3) Close-up shows repeated interlocking compliant cells and black conductive junctions; scale bar 1 cm. (b1-b3) Grapes, an apple, and a kumquat are placed on the matrix while a phone displays the predicted class. Reported recognition accuracies are 92.7 percent for grape, 96.0 percent for apple, and 77.6 percent for kumquat.}
    \label{Application4}
\end{figure}

To validate this potential, we developed an 8-channel tactile sensing matrix, synchronizing two sensing hardware units to acquire orthogonal 4-channel X and 4-channel Y data. When an object is placed on the matrix, its contact area, shape, and weight leave a distinct deformation distribution pattern across the structure. We used three fruits of varying size and weight—grape, apple, and kumquat—as proxy objects to systematically cover a range of contact sizes and pressure distributions. A lightweight machine learning model classifies these deformation signatures, achieving recognition accuracies of 92.7\%, 96.0\%, and 77.6\% respectively.

This application demonstrates that an X-Hinges matrix can effectively distinguish objects with different physical characteristics. Its printable, customizable nature makes it a promising building block for robotic electronic skin and intelligent sensing surfaces—extending tactile perception to physical surfaces of arbitrary shape at minimal fabrication cost.

\section{EVALUATION}

\subsection{Selectivity of Individual Sensing Configurations}

\begin{figure}[h]
    \includegraphics[width=0.96\linewidth]{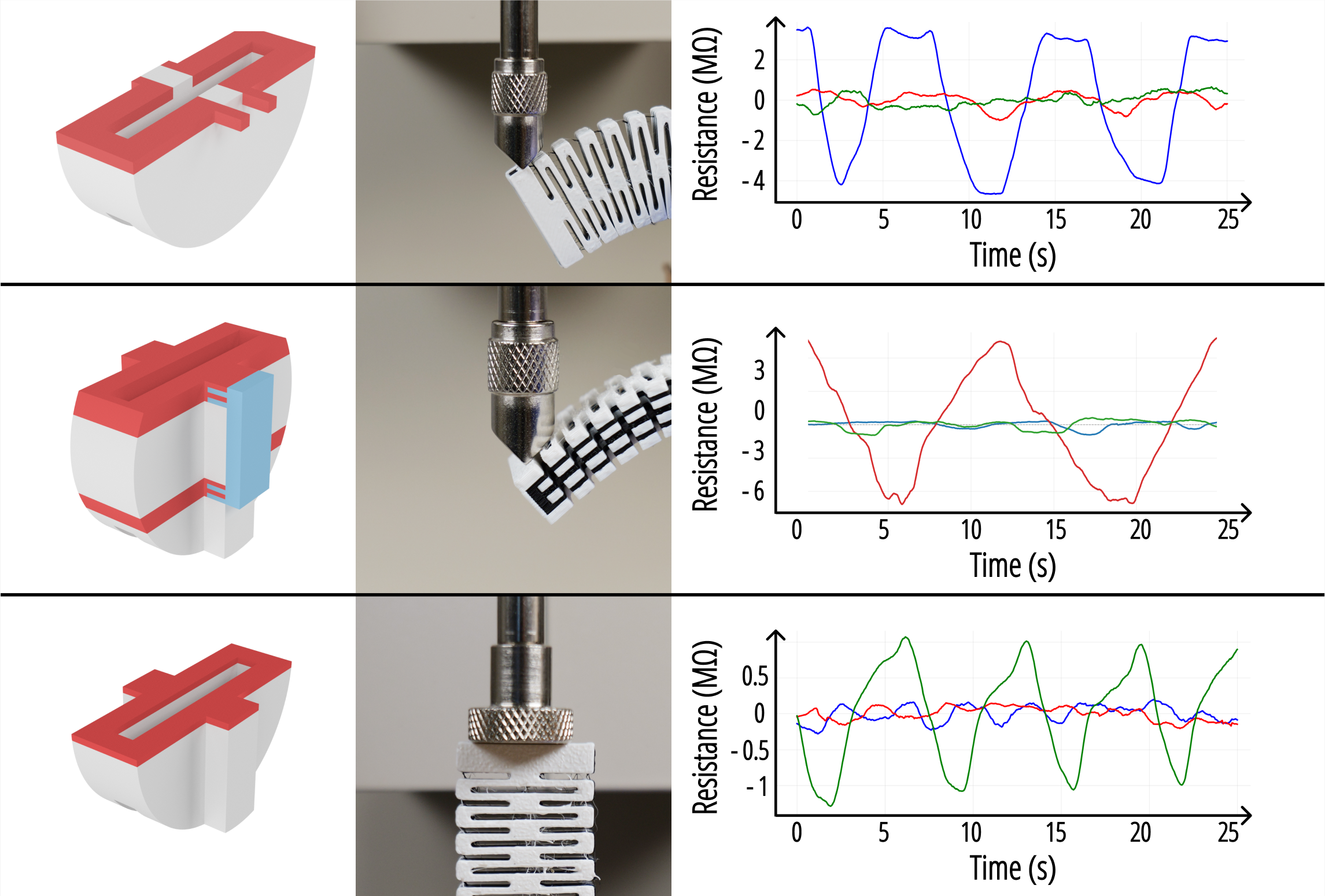}
    \caption{Normalized differential resistance responses for the three configurations under lateral (blue), vertical (red), and axial (green) motions.}
    \Description{Selectivity tests for three individual sensing configurations, arranged in rows. Each row contains a rendering of the sensing-element geometry, a photograph of a universal-testing-machine probe deforming the printed specimen, and a resistance-versus-time plot. The top lateral configuration responds strongly in blue during lateral bending, swinging roughly from +3.5 to -4.5 megaohms while the red and green off-axis responses remain near zero. The middle vertical configuration responds strongly in red, roughly +5 to -6.5 megaohms, with much smaller blue and green changes. The bottom axial configuration responds primarily in green, roughly -1.3 to +0.9 megaohms, while blue and red remain comparatively small.}
    \label{cross_sensitivity_evaluation}
\end{figure}

We evaluate the motion selectivity of the three sensing configurations introduced in Section~\ref{sensing}. Three rectangular X-Hinge specimens ($40 \times 20 \times 10$\,mm) were fabricated, each embedding a single distinct configuration. Using a universal testing machine to enforce consistent displacement, each specimen was subjected to controlled deformations along three axes—lateral bending ($45^\circ$), vertical bending ($60^\circ$), and axial displacement ($30$\,mm). Resistance responses were recorded synchronously across all three axes (Figure~\ref{cross_sensitivity_evaluation}).

\begin{table}[h]
    \caption{Normalized crosstalk matrix for sensing configurations under single-axis actuation.}
    \Description{Normalized crosstalk matrix for three sensing configurations under single-axis actuation. Columns are applied lateral motion at 45 degrees, vertical motion at 60 degrees, and axial motion at 30 mm. Lateral configuration responses are 100 percent, 13.9 percent, and 12.8 percent. Vertical configuration responses are 8.2 percent, 100 percent, and 10.5 percent. Axial configuration responses are 18.5 percent, 19.2 percent, and 100 percent. Each configuration has its maximum response on the intended-motion diagonal; off-axis responses range from 8.2 to 19.2 percent.}
    \label{tab:crosstalk} 
    \small
    \centering
    \begin{tabular}{lccc}
    \toprule
    \textbf{Sensing} & \multicolumn{3}{c}{\textbf{Applied Motion}} \\
    \cmidrule(lr){2-4}
    \textbf{Configuration} & \textbf{Lateral ($45^\circ$)} & \textbf{Vertical
    ($60^\circ$)} & \textbf{Axial ($30$\,mm)} \\
    \midrule
    Lateral Config  & \textbf{100\%} & 13.9\% & 12.8\% \\
    Vertical Config & 8.2\% & \textbf{100\%} & 10.5\% \\
    Axial Config    & 18.5\% & 19.2\% & \textbf{100\%} \\
\bottomrule
\end{tabular}
\end{table}

All three configurations showed selectivity for their intended motions (Table~\ref{tab:crosstalk}), with limited off-axis coupling in the lateral and vertical configurations. The axial configuration exhibited moderately greater coupling, likely because axial loading induced parasitic bending that activated the other resistance bridges. This coupling persisted after co-integration and is further examined in Section~\ref{evaluation_integrated}.

\subsection{Multi-Axis Motion Sensing in the Integrated Structure}\label{evaluation_integrated}

\begin{figure}[h]
    \includegraphics[width=0.96\linewidth]{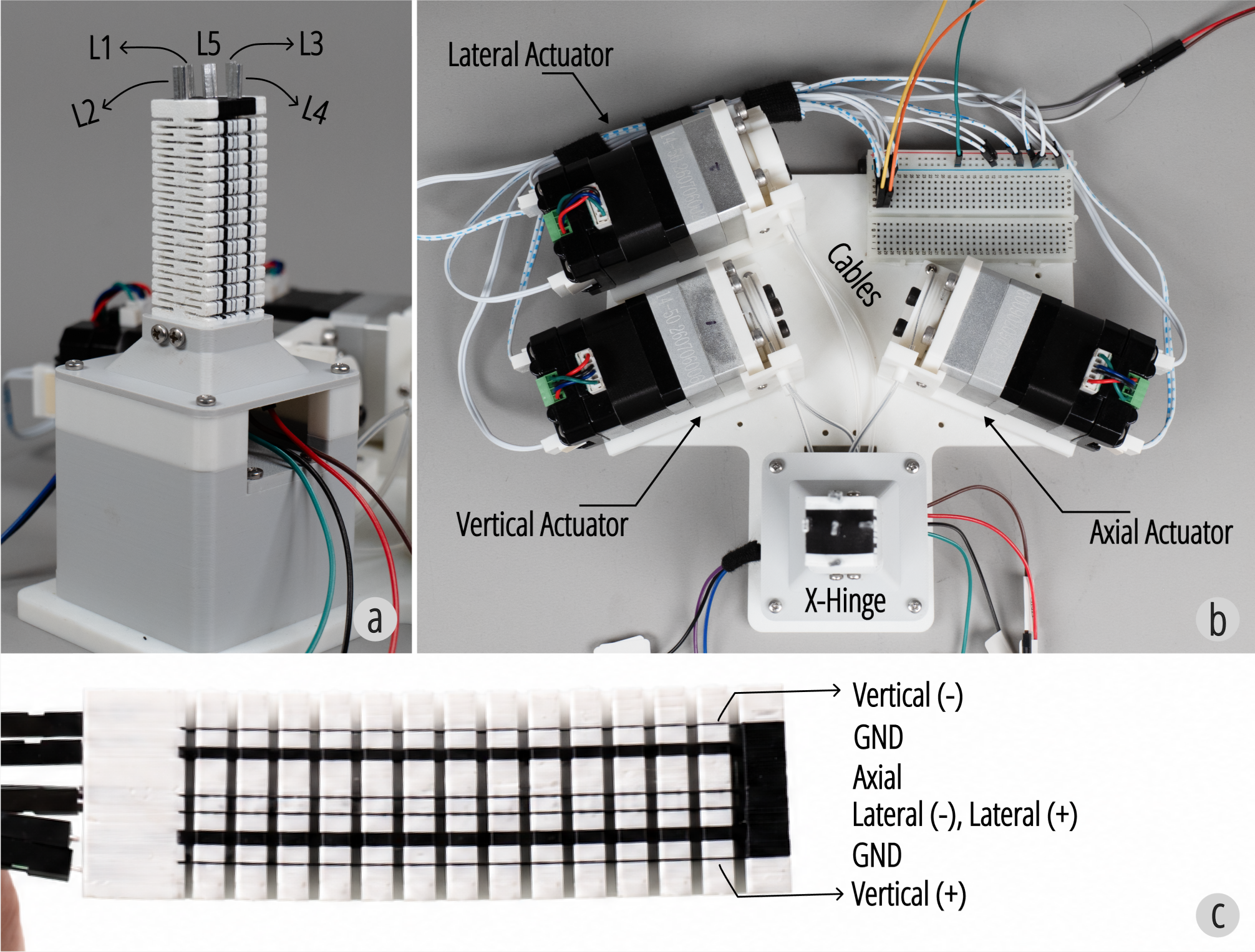}
    \caption{Cable-driven setup for multi-axis evaluation. (a) Five-cable actuation. (b) Three-axis test rig. (c) Integrated X-Hinges specimen.}
    \Description{Cable-driven three-axis evaluation setup. (a) A tall integrated X-Hinges specimen is fixed vertically on a test base; five cables labeled L1 through L5 attach around its top to produce controlled motion. (b) Top view of the test rig shows three motorized units labeled lateral actuator, vertical actuator, and axial actuator, with their cables converging on the central X-Hinge. (c) Close-up of the integrated specimen identifies seven conductive lines from top to bottom: vertical negative, ground, axial, lateral negative and lateral positive, ground, and vertical positive.}
    \label{cable-driven-test}
\end{figure}
\begin{figure}[h]
    \includegraphics[width=0.96\linewidth]{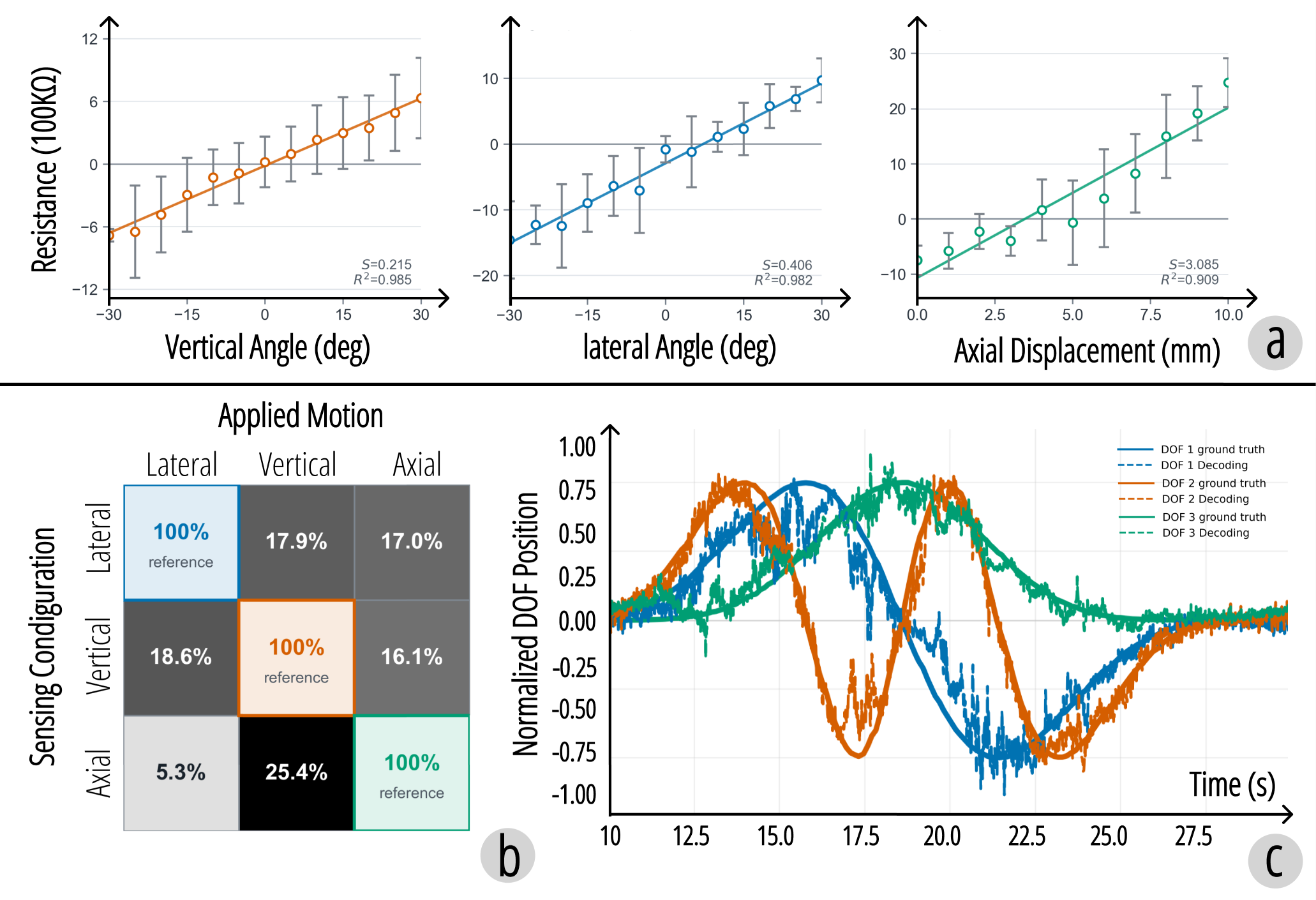}
    \caption{Three-DOF sensing performance. (a) Resistance–motion response. (b) Cross-talk matrix. (c) Motion decoding.}
    \Description{Three-DOF sensing performance. (a) Three calibration plots show approximately linear resistance-motion relationships with error bars: vertical angle from -30 to +30 degrees has slope 0.215 and R-squared 0.985; lateral angle over the same range has slope 0.406 and R-squared 0.982; axial displacement from 0 to 10 mm has slope 3.085 and R-squared 0.909. (b) The normalized crosstalk matrix has 100 percent diagonal responses. Off-axis values are 17.9 and 17.0 percent for the lateral configuration, 18.6 and 16.1 percent for the vertical configuration, and 5.3 and 25.4 percent for the axial configuration. (c) Solid ground-truth and dashed decoded trajectories for all three DOFs broadly overlap over time, with mean absolute errors of 7.35 degrees lateral, 6.59 degrees vertical, and 1.35 mm axial.}
    \label{Evaluation_8_2_2}
\end{figure}

\red{
We next evaluate multi-axis motion sensing in the integrated structure. Using the cable-driven setup in Figure~\ref{cable-driven-test}, we first actuate the specimen independently along the lateral, vertical, and axial degrees of freedom to characterize its resistance--motion relationships and residual crosstalk. As shown in Figure~\ref{Evaluation_8_2_2}b, each sensing configuration exhibits the strongest response to its intended motion, while the off-axis responses are 17.9\% for lateral sensing, 18.6\% for vertical sensing, and 25.4\% for axial sensing.
}

\red{
We then evaluate the integrated structure under coupled multi-axis actuation. Figure~\ref{Evaluation_8_2_2}c compares the reference motion trajectories derived from the motor encoder readings with those decoded from the resistance signals using the TCN. For the lateral, vertical, and axial degrees of freedom, respectively, the TCN achieves mean absolute errors of $7.35^\circ$, $6.59^\circ$, and $1.35\,\mathrm{mm}$.
}
\subsection{Self-Calibration Across Prints and Temporal Drift}\label{evaluation_self_calibration}

\red{
We evaluate the self-calibration workflow introduced in Section~\ref{self-calibration} by examining print-to-print variability and temporal drift in single-DOF X-Hinges specimens. We first trained a base TCN motion decoder on an original specimen (S1) and fabricated a second specimen of the same geometry (S2) to assess model transfer across prints. To induce pronounced drift in S1, we subjected it to 1,000 loading cycles over 20 hours, followed by three additional days of water immersion as an accelerated stress protocol. We then re-evaluated S1 to quantify the resulting degradation in decoding performance.
}

\begin{figure}[h]
    \includegraphics[width=0.96\linewidth]{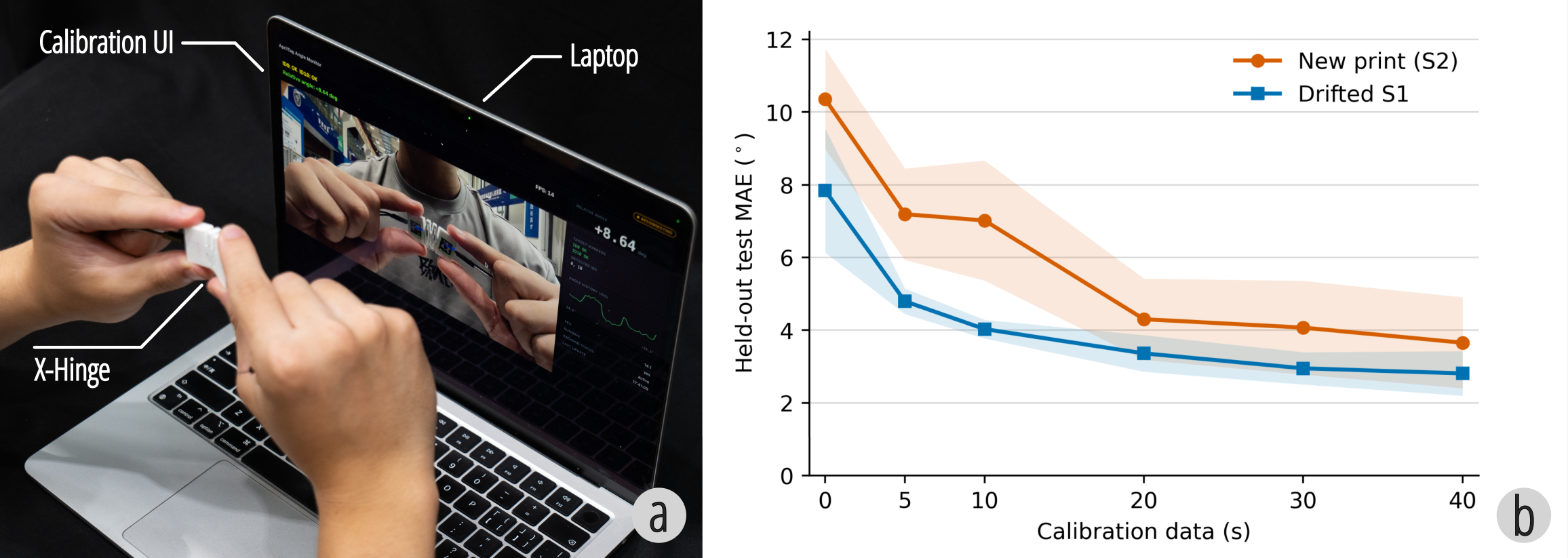}
    \caption{Self-calibration setup and performance: (a) calibration interface and (b) held-out test MAE versus calibration-data duration for S2 and drifted S1}
    \Description{Self-calibration setup and results. (a) A user bends a small X-Hinge in front of a laptop camera. The laptop's calibration interface tracks the printed markers and displays the measured angle and resistance history. (b) Held-out test mean absolute error decreases as labeled calibration duration increases from 0 to 40 seconds. The new print S2, shown in orange circles, drops from about 10 degrees to under 4 degrees; drifted S1, shown in blue squares, drops from about 8 degrees to about 3 degrees. Shaded bands indicate variability, which also narrows with more calibration data.}
    \label{calibration_evaluation}
\end{figure}

\red{
We then applied specimen-specific calibration to both the newly printed S2 and the drifted S1 (Figure~\ref{calibration_evaluation}a). As shown in Figure~\ref{calibration_evaluation}b, directly applying the S1-trained model yields held-out MAEs of $10.02^\circ$ on S2 and $8.20^\circ$ on drifted S1. With labeled calibration data, the errors decrease to $3.96^\circ$ and $3.71^\circ$, respectively.
}

\subsection{Sensing Element–Trace Interface Characterization}\label{evaluation_4}
X-Hinges integrates two conductive filaments whose resistivities differ by three orders of magnitude ($\sim\!1.23\times10^{4}\,\Omega\cdot\mathrm{cm}$ vs.\ $\sim\!8.75\,\Omega\cdot\mathrm{cm}$). The interface between them introduces a series resistance $r_i$ that can mask deformation-induced resistance changes. We evaluate four interface geometries to minimize this junction resistance: (1)~\textbf{Parallel Beam}—filaments run side-by-side (\(5.6\,\mathrm{mm}\) contact length); (2)~\textbf{Direct Contact}—end-to-end at a flat face (\(1.2\,\mathrm{mm}\)); (3)~\textbf{Cross Beam}—orthogonal crossing (\(5.2\,\mathrm{mm}\) effective length); (4)~\textbf{Layer Overlap}—one material deposited onto the other across print layers (\(1.2\,\mathrm{mm}\) overlap).

\begin{figure}[h]
    \includegraphics[width=0.96\linewidth]{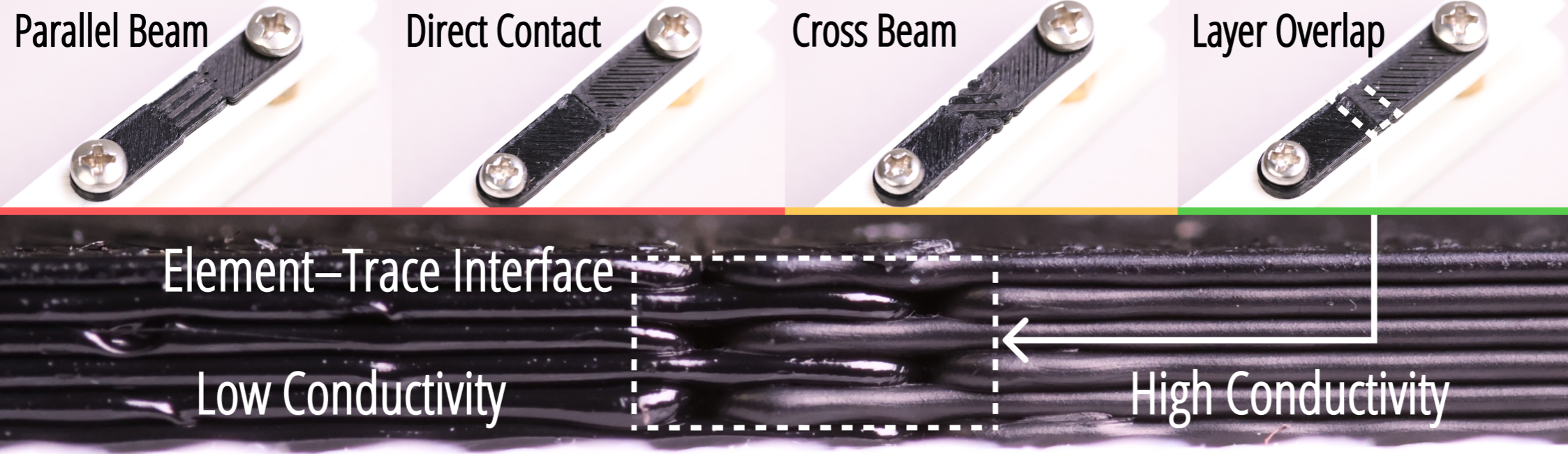}
    \caption{Four interface structures for connecting high- and low-resistivity filaments.}
    \Description{Four printed geometries for joining the high-resistivity sensing element to the low-resistivity trace: parallel beam, direct end-to-end contact, orthogonal cross beam, and layer overlap. Photographs across the top show each black test strip clamped between two screws. A magnified side view below highlights the element-trace interface, with the low-conductivity material on the left and high-conductivity material on the right. A dashed rectangle and arrow identify the stacked contact region used by the layer-overlap geometry.}
    \label{interface}
\end{figure}

Table~\ref{tab:interface} shows net interface resistance for each geometry (5 replicates, undeformed state). Layer Overlap achieves the lowest resistance (\(22.3\,\mathrm{k}\Omega\)) and best consistency (RSD \(8.1\%\)), approximately $6\times$ lower than Cross Beam and $43\times$ lower than Direct Contact.

\begin{table}[h]
\caption{Net interface resistance ($r_i$) after subtracting \(241.44\,\mathrm{k}\Omega\) baseline. RSD: relative standard deviation.}
\Description{Net element-trace interface resistance after subtracting a 241.44 kiloohm baseline; each geometry has five samples. Parallel Beam has mean 1697.4 kiloohms, standard deviation 424.9 kiloohms, and relative standard deviation 25.0 percent. Direct Contact has mean 952.2 kiloohms, standard deviation 178.4 kiloohms, and relative standard deviation 18.7 percent. Cross Beam has mean 132.8 kiloohms, standard deviation 17.2 kiloohms, and relative standard deviation 13.0 percent. Layer Overlap has the lowest and most consistent resistance: mean 22.3 kiloohms, standard deviation 1.8 kiloohms, and relative standard deviation 8.1 percent.}
\label{tab:interface}
\small
\begin{tabular}{lcccc}
\hline
\textbf{Geometry} & \textbf{$N$} & \textbf{Mean ($k\Omega$)} &
\textbf{Std ($k\Omega$)} & \textbf{RSD (\%)} \\
\hline
Parallel Beam  & 5 & 1697.4 & 424.9 & 25.0 \\
Direct Contact & 5 &  952.2 & 178.4 & 18.7 \\
Cross Beam     & 5 &  132.8 &  17.2 & 13.0 \\
Layer Overlap  & 5 &   22.3 &   1.8 & 8.1 \\
\hline
\end{tabular}
\end{table}

\section{LIMITATIONS AND FUTURE WORK}
\subsection{Material Properties and Sensing Precision}
Although X-Hinges \red{mitigate multi-DOF coupling} through strategic material configurations and TCN-based hysteresis mitigation, conductive TPU filaments \red{still exhibit} hysteresis and viscoelastic creep, \red{limiting high-
precision motion tracking} during rapid or repetitive deformations.

Future work \red{will investigate resistance-change mechanisms and microscopic contact between high- and low- conductivity phases} within multi-material 3D-printed interfaces \red{to develop robust models and fabrication strategies that improve sensing stability and reduce reliance on data-driven compensation.}

\subsection{Instance Variability and Calibration}
Inherent print and material variability limits model transfer across X-Hinge instances. Although vision-based calibration enables instance-specific adaptation, it requires external tracking. \red{This also complicates fully integrated three-DOF evaluation, since passive structures typically lack encoders or actuators for synchronized ground-truth collection.} Future work will explore intrinsic resistance signatures for self-supervised, reference-free calibration toward a truly ``plug-and-play'' workflow.

\subsection{\red{Generalizability and Practical Adoption}}
\red{Our controlled study demonstrates three-axis motion estimation in one prototype, but does not fully validate three-DoF mechanisms across geometries, stiffnesses, or prints. Requiring multi-material FDM, dedicated electronics, and calibration, X-Hinges primarily targets fabrication researchers, interaction designers, and engineers. As a research prototyping method, X-Hinges reduces post-fabrication sensor integration. Future work will broaden validation across geometries, stiffness configurations, and unconstrained interactions; simplify the sensing hardware and calibration workflow.}
\section{CONCLUSION}

In this paper, we present X-Hinges, a design and fabrication method that co-fabricates sensing and structure as a unified whole through multi-material FDM 3D printing. By integrating differential measurement principles directly into printed physical geometry, X-Hinges \red{reduces cross-axis coupling} of multi-DOF deformations—a capability that single-material approaches have not been able to provide. Built upon a versatile design space spanning motion primitives, mechanical stiffness, and geometric forms, and supported by a custom high-precision sensing pipeline and an interactive design tool, X-Hinges enables designers to rapidly prototype personalized interactive artifacts with continuous, multi-axis self-sensing embedded from the first print.

Through diverse applications—from a teleoperation glove and a tangible game controller to a deployable interactive lamp and a tactile sensing matrix—we demonstrate that X-Hinges functions as a general-purpose building block for functionally integrated interactive objects. \red{Technical evaluations characterize sensing-configuration selectivity, multi-axis motion-estimation errors, and self-calibration performance.} We hope X-Hinges inspires new ways of conceiving interactive objects as self-aware structures in which form, function, and sensing are unified through fabrication.

\begin{acks}
We thank Dingning Cao for visual design support, and Chengdu Changshu Robot Co., Ltd. for providing the OpenArmX platform.
\end{acks}

\bibliographystyle{ACM-Reference-Format}
\bibliography{cite}


\begin{thebibliography}{40}


\ifx \showCODEN    \undefined \def \showCODEN     #1{\unskip}     \fi
\ifx \showISBNx    \undefined \def \showISBNx     #1{\unskip}     \fi
\ifx \showISBNxiii \undefined \def \showISBNxiii  #1{\unskip}     \fi
\ifx \showISSN     \undefined \def \showISSN      #1{\unskip}     \fi
\ifx \showLCCN     \undefined \def \showLCCN      #1{\unskip}     \fi
\ifx \shownote     \undefined \def \shownote      #1{#1}          \fi
\ifx \showarticletitle \undefined \def \showarticletitle #1{#1}   \fi
\ifx \showURL      \undefined \def \showURL       {\relax}        \fi
\providecommand\bibfield[2]{#2}
\providecommand\bibinfo[2]{#2}
\providecommand\natexlab[1]{#1}
\providecommand\showeprint[2][]{arXiv:#2}

\bibitem[B\"{a}cher et~al\mbox{.}(2016)]%
        {10.1145/2858036.2858354}
\bibfield{author}{\bibinfo{person}{Moritz B\"{a}cher},
  \bibinfo{person}{Benjamin Hepp}, \bibinfo{person}{Fabrizio Pece},
  \bibinfo{person}{Paul~G. Kry}, \bibinfo{person}{Bernd Bickel},
  \bibinfo{person}{Bernhard Thomaszewski}, {and} \bibinfo{person}{Otmar
  Hilliges}.} \bibinfo{year}{2016}\natexlab{}.
\newblock \showarticletitle{DefSense: Computational Design of Customized
  Deformable Input Devices}. In \bibinfo{booktitle}{\emph{Proceedings of the
  2016 CHI Conference on Human Factors in Computing Systems}} (San Jose,
  California, USA) \emph{(\bibinfo{series}{CHI '16})}.
  \bibinfo{publisher}{Association for Computing Machinery},
  \bibinfo{address}{New York, NY, USA}, \bibinfo{pages}{3806–3816}.
\newblock
\showISBNx{9781450333627}
\href{https://doi.org/10.1145/2858036.2858354}{doi:\nolinkurl{10.1145/2858036.2858354}}


\bibitem[Bae et~al\mbox{.}(2025)]%
        {bae2025computational}
\bibfield{author}{\bibinfo{person}{S~Sandra Bae}, \bibinfo{person}{Takanori
  Fujiwara}, \bibinfo{person}{Danielle~Albers Szafir}, \bibinfo{person}{Ellen
  Yi-Luen Do}, {and} \bibinfo{person}{Michael~L Rivera}.}
  \bibinfo{year}{2025}\natexlab{}.
\newblock \showarticletitle{Computational Design and Single-Wire Sensing of 3D
  Printed Objects with Integrated Capacitive Touchpoints}.
\newblock \bibinfo{journal}{\emph{arXiv preprint arXiv:2509.25387}}
  (\bibinfo{year}{2025}).
\newblock


\bibitem[Bae et~al\mbox{.}(2024)]%
        {network}
\bibfield{author}{\bibinfo{person}{S.~Sandra Bae}, \bibinfo{person}{Takanori
  Fujiwara}, \bibinfo{person}{Anders Ynnerman}, \bibinfo{person}{Ellen Yi-Luen
  Do}, \bibinfo{person}{Michael~L. Rivera}, {and}
  \bibinfo{person}{Danielle~Albers Szafir}.} \bibinfo{year}{2024}\natexlab{}.
\newblock \showarticletitle{A Computational Design Pipeline to Fabricate
  Sensing Network Physicalizations}.
\newblock \bibinfo{journal}{\emph{IEEE Transactions on Visualization and
  Computer Graphics}} \bibinfo{volume}{30}, \bibinfo{number}{1}
  (\bibinfo{year}{2024}), \bibinfo{pages}{913--923}.
\newblock
\href{https://doi.org/10.1109/TVCG.2023.3327198}{doi:\nolinkurl{10.1109/TVCG.2023.3327198}}


\bibitem[Bai et~al\mbox{.}(2018)]%
        {bai2018empirical}
\bibfield{author}{\bibinfo{person}{Shaojie Bai}, \bibinfo{person}{J~Zico
  Kolter}, {and} \bibinfo{person}{Vladlen Koltun}.}
  \bibinfo{year}{2018}\natexlab{}.
\newblock \showarticletitle{An empirical evaluation of generic convolutional
  and recurrent networks for sequence modeling}.
\newblock \bibinfo{journal}{\emph{arXiv preprint arXiv:1803.01271}}
  (\bibinfo{year}{2018}).
\newblock


\bibitem[Boem and Troiano(2019)]%
        {10.1145/3322276.3322347}
\bibfield{author}{\bibinfo{person}{Alberto Boem} {and}
  \bibinfo{person}{Giovanni~Maria Troiano}.} \bibinfo{year}{2019}\natexlab{}.
\newblock \showarticletitle{Non-Rigid HCI: A Review of Deformable Interfaces
  and Input}. In \bibinfo{booktitle}{\emph{Proceedings of the 2019 on Designing
  Interactive Systems Conference}} (San Diego, CA, USA)
  \emph{(\bibinfo{series}{DIS '19})}. \bibinfo{publisher}{Association for
  Computing Machinery}, \bibinfo{address}{New York, NY, USA},
  \bibinfo{pages}{885–906}.
\newblock
\showISBNx{9781450358507}
\href{https://doi.org/10.1145/3322276.3322347}{doi:\nolinkurl{10.1145/3322276.3322347}}


\bibitem[BURNS(1965)]%
        {burns1965kinetostatic}
\bibfield{author}{\bibinfo{person}{RICHARD~HOLLINGTON BURNS}.}
  \bibinfo{year}{1965}\natexlab{}.
\newblock \bibinfo{booktitle}{\emph{The kinetostatic synthesis of flexible link
  mechanisms}}.
\newblock \bibinfo{publisher}{Yale University}.
\newblock


\bibitem[Burstyn et~al\mbox{.}(2015)]%
        {PrintPut}
\bibfield{author}{\bibinfo{person}{Jesse Burstyn}, \bibinfo{person}{Nicholas
  Fellion}, \bibinfo{person}{Paul Strohmeier}, {and} \bibinfo{person}{Roel
  Vertegaal}.} \bibinfo{year}{2015}\natexlab{}.
\newblock \showarticletitle{PrintPut: Resistive and Capacitive Input Widgets
  for Interactive 3D Prints}. In \bibinfo{booktitle}{\emph{Human-Computer
  Interaction – INTERACT 2015}}. \bibinfo{publisher}{Springer-Verlag},
  \bibinfo{address}{Berlin, Heidelberg}, \bibinfo{pages}{332–339}.
\newblock
\showISBNx{978-3-319-22700-9}
\href{https://doi.org/10.1007/978-3-319-22701-6_25}{doi:\nolinkurl{10.1007/978-3-319-22701-6_25}}


\bibitem[Dogan et~al\mbox{.}(2021)]%
        {10.1145/3472749.3474733}
\bibfield{author}{\bibinfo{person}{Mustafa~Doga Dogan},
  \bibinfo{person}{Steven~Vidal Acevedo~Colon}, \bibinfo{person}{Varnika
  Sinha}, \bibinfo{person}{Kaan Ak\c{s}it}, {and} \bibinfo{person}{Stefanie
  Mueller}.} \bibinfo{year}{2021}\natexlab{}.
\newblock \showarticletitle{SensiCut: Material-Aware Laser Cutting Using
  Speckle Sensing and Deep Learning}. In \bibinfo{booktitle}{\emph{The 34th
  Annual ACM Symposium on User Interface Software and Technology}} (Virtual
  Event, USA) \emph{(\bibinfo{series}{UIST '21})}.
  \bibinfo{publisher}{Association for Computing Machinery},
  \bibinfo{address}{New York, NY, USA}, \bibinfo{pages}{24–38}.
\newblock
\showISBNx{9781450386357}
\href{https://doi.org/10.1145/3472749.3474733}{doi:\nolinkurl{10.1145/3472749.3474733}}


\bibitem[Dogan et~al\mbox{.}(2022)]%
        {10.1145/3491102.3501951}
\bibfield{author}{\bibinfo{person}{Mustafa~Doga Dogan}, \bibinfo{person}{Ahmad
  Taka}, \bibinfo{person}{Michael Lu}, \bibinfo{person}{Yunyi Zhu},
  \bibinfo{person}{Akshat Kumar}, \bibinfo{person}{Aakar Gupta}, {and}
  \bibinfo{person}{Stefanie Mueller}.} \bibinfo{year}{2022}\natexlab{}.
\newblock \showarticletitle{InfraredTags: Embedding Invisible AR Markers and
  Barcodes Using Low-Cost, Infrared-Based 3D Printing and Imaging Tools}. In
  \bibinfo{booktitle}{\emph{Proceedings of the 2022 CHI Conference on Human
  Factors in Computing Systems}} (New Orleans, LA, USA)
  \emph{(\bibinfo{series}{CHI '22})}. \bibinfo{publisher}{Association for
  Computing Machinery}, \bibinfo{address}{New York, NY, USA}, Article
  \bibinfo{articleno}{269}, \bibinfo{numpages}{12}~pages.
\newblock
\showISBNx{9781450391573}
\href{https://doi.org/10.1145/3491102.3501951}{doi:\nolinkurl{10.1145/3491102.3501951}}


\bibitem[Eduardo Aguilar-Segovia et~al\mbox{.}(2025)]%
        {eduardo_aguilar-segovia_parametric_2025}
\bibfield{author}{\bibinfo{person}{José Eduardo Aguilar-Segovia},
  \bibinfo{person}{Fabien Grzeskowiak}, \bibinfo{person}{Sylvain Lefebvre},
  \bibinfo{person}{Marie Babel}, {and} \bibinfo{person}{Sylvain Guégan}.}
  \bibinfo{year}{2025}\natexlab{}.
\newblock \showarticletitle{Parametric {Foam}-{Like} {Capacitive} {Sensors} for
  3-{D} {Printing} of {Deformable} {Parts} {With} {Sensing} {Capabilities}}.
\newblock \bibinfo{journal}{\emph{IEEE Sensors Journal}} \bibinfo{volume}{25},
  \bibinfo{number}{3} (\bibinfo{date}{Feb.} \bibinfo{year}{2025}),
  \bibinfo{pages}{4261--4272}.
\newblock
\showISSN{1558-1748}
\href{https://doi.org/10.1109/JSEN.2024.3510138}{doi:\nolinkurl{10.1109/JSEN.2024.3510138}}


\bibitem[Gong et~al\mbox{.}(2021)]%
        {10.1145/3472749.3474806}
\bibfield{author}{\bibinfo{person}{Jun Gong}, \bibinfo{person}{Olivia Seow},
  \bibinfo{person}{Cedric Honnet}, \bibinfo{person}{Jack Forman}, {and}
  \bibinfo{person}{Stefanie Mueller}.} \bibinfo{year}{2021}\natexlab{}.
\newblock \showarticletitle{MetaSense: Integrating Sensing Capabilities into
  Mechanical Metamaterial}. In \bibinfo{booktitle}{\emph{The 34th Annual ACM
  Symposium on User Interface Software and Technology}} (Virtual Event, USA)
  \emph{(\bibinfo{series}{UIST '21})}. \bibinfo{publisher}{Association for
  Computing Machinery}, \bibinfo{address}{New York, NY, USA},
  \bibinfo{pages}{1063–1073}.
\newblock
\showISBNx{9781450386357}
\href{https://doi.org/10.1145/3472749.3474806}{doi:\nolinkurl{10.1145/3472749.3474806}}


\bibitem[Hoffmann(1974)]%
        {hoffmann1974applying}
\bibfield{author}{\bibinfo{person}{Karl Hoffmann}.}
  \bibinfo{year}{1974}\natexlab{}.
\newblock \bibinfo{booktitle}{\emph{Applying the wheatstone bridge circuit}}.
\newblock \bibinfo{publisher}{HBM Darmstadt, Germany}.
\newblock


\bibitem[Howell(2013)]%
        {howell2013compliant}
\bibfield{author}{\bibinfo{person}{Larry~L Howell}.}
  \bibinfo{year}{2013}\natexlab{}.
\newblock \showarticletitle{Compliant mechanisms}. In
  \bibinfo{booktitle}{\emph{21st century kinematics: The 2012 NSF Workshop}}.
  Springer, \bibinfo{pages}{189--216}.
\newblock


\bibitem[Ishii and Ullmer(1997)]%
        {10.1145/258549.258715}
\bibfield{author}{\bibinfo{person}{Hiroshi Ishii} {and} \bibinfo{person}{Brygg
  Ullmer}.} \bibinfo{year}{1997}\natexlab{}.
\newblock \showarticletitle{Tangible bits: towards seamless interfaces between
  people, bits and atoms}. In \bibinfo{booktitle}{\emph{Proceedings of the ACM
  SIGCHI Conference on Human Factors in Computing Systems}} (Atlanta, Georgia,
  USA) \emph{(\bibinfo{series}{CHI '97})}. \bibinfo{publisher}{Association for
  Computing Machinery}, \bibinfo{address}{New York, NY, USA},
  \bibinfo{pages}{234–241}.
\newblock
\showISBNx{0897918029}
\href{https://doi.org/10.1145/258549.258715}{doi:\nolinkurl{10.1145/258549.258715}}


\bibitem[Jacobsen et~al\mbox{.}(2009)]%
        {JACOBSEN20092098}
\bibfield{author}{\bibinfo{person}{Joseph~O. Jacobsen}, \bibinfo{person}{Guimin
  Chen}, \bibinfo{person}{Larry~L. Howell}, {and} \bibinfo{person}{Spencer~P.
  Magleby}.} \bibinfo{year}{2009}\natexlab{}.
\newblock \showarticletitle{Lamina Emergent Torsional (LET) Joint}.
\newblock \bibinfo{journal}{\emph{Mechanism and Machine Theory}}
  \bibinfo{volume}{44}, \bibinfo{number}{11} (\bibinfo{year}{2009}),
  \bibinfo{pages}{2098--2109}.
\newblock
\showISSN{0094-114X}
\href{https://doi.org/10.1016/j.mechmachtheory.2009.05.015}{doi:\nolinkurl{10.1016/j.mechmachtheory.2009.05.015}}


\bibitem[Kota et~al\mbox{.}(2001)]%
        {kota2001design}
\bibfield{author}{\bibinfo{person}{Sridhar Kota}, \bibinfo{person}{Jinyong
  Joo}, \bibinfo{person}{Zhe Li}, \bibinfo{person}{Steven~M Rodgers}, {and}
  \bibinfo{person}{Jeff Sniegowski}.} \bibinfo{year}{2001}\natexlab{}.
\newblock \showarticletitle{Design of compliant mechanisms: applications to
  MEMS}.
\newblock \bibinfo{journal}{\emph{Analog integrated circuits and signal
  processing}} \bibinfo{volume}{29}, \bibinfo{number}{1}
  (\bibinfo{year}{2001}), \bibinfo{pages}{7--15}.
\newblock


\bibitem[Li et~al\mbox{.}(2026b)]%
        {Y-zipper}
\bibfield{author}{\bibinfo{person}{Jiaji Li}, \bibinfo{person}{Xiang Chang},
  \bibinfo{person}{Mingming Li}, \bibinfo{person}{Dingning Cao},
  \bibinfo{person}{Maxine Perroni-Scharf}, \bibinfo{person}{Jeremy Mrzyglocki},
  \bibinfo{person}{Takumi Yamamoto}, \bibinfo{person}{William Freeman}, {and}
  \bibinfo{person}{Stefanie Mueller}.} \bibinfo{year}{2026}\natexlab{b}.
\newblock \showarticletitle{Y-zipper: 3D Printing Flexible–Rigid Transition
  Mechanism for Rapid and Reversible Assembly}. In
  \bibinfo{booktitle}{\emph{Proceedings of the 2026 CHI Conference on Human
  Factors in Computing Systems}} \emph{(\bibinfo{series}{CHI '26})}.
  \bibinfo{publisher}{Association for Computing Machinery},
  \bibinfo{address}{New York, NY, USA}, Article \bibinfo{articleno}{754},
  \bibinfo{numpages}{17}~pages.
\newblock
\showISBNx{9798400722783}
\href{https://doi.org/10.1145/3772318.3790723}{doi:\nolinkurl{10.1145/3772318.3790723}}


\bibitem[Li et~al\mbox{.}(2025a)]%
        {Xstrings}
\bibfield{author}{\bibinfo{person}{Jiaji Li}, \bibinfo{person}{Shuyue Feng},
  \bibinfo{person}{Maxine Perroni-Scharf}, \bibinfo{person}{Yujia Liu},
  \bibinfo{person}{Emily Guan}, \bibinfo{person}{Guanyun Wang}, {and}
  \bibinfo{person}{Stefanie Mueller}.} \bibinfo{year}{2025}\natexlab{a}.
\newblock \showarticletitle{Xstrings: 3D Printing Cable-Driven Mechanism for
  Actuation, Deformation, and Manipulation}. In
  \bibinfo{booktitle}{\emph{Proceedings of the 2025 CHI Conference on Human
  Factors in Computing Systems}} \emph{(\bibinfo{series}{CHI '25})}.
  \bibinfo{publisher}{Association for Computing Machinery},
  \bibinfo{address}{New York, NY, USA}, Article \bibinfo{articleno}{6},
  \bibinfo{numpages}{17}~pages.
\newblock
\showISBNx{9798400713941}
\href{https://doi.org/10.1145/3706598.3714282}{doi:\nolinkurl{10.1145/3706598.3714282}}


\bibitem[Li et~al\mbox{.}(2023)]%
        {all-in-one}
\bibfield{author}{\bibinfo{person}{Jiaji Li}, \bibinfo{person}{Mingming Li},
  \bibinfo{person}{Junzhe Ji}, \bibinfo{person}{Deying Pan},
  \bibinfo{person}{Yitao Fan}, \bibinfo{person}{Kuangqi Zhu},
  \bibinfo{person}{Yue Yang}, \bibinfo{person}{Zihan Yan},
  \bibinfo{person}{Lingyun Sun}, \bibinfo{person}{Ye Tao}, {and}
  \bibinfo{person}{Guanyun Wang}.} \bibinfo{year}{2023}\natexlab{}.
\newblock \showarticletitle{All-in-One Print: Designing and 3D Printing Dynamic
  Objects Using Kinematic Mechanism Without Assembly}. In
  \bibinfo{booktitle}{\emph{Proceedings of the 2023 CHI Conference on Human
  Factors in Computing Systems}} (Hamburg, Germany) \emph{(\bibinfo{series}{CHI
  '23})}. \bibinfo{publisher}{Association for Computing Machinery},
  \bibinfo{address}{New York, NY, USA}, Article \bibinfo{articleno}{689},
  \bibinfo{numpages}{15}~pages.
\newblock
\showISBNx{9781450394215}
\href{https://doi.org/10.1145/3544548.3581440}{doi:\nolinkurl{10.1145/3544548.3581440}}


\bibitem[Li et~al\mbox{.}(2026a)]%
        {10.1145/3772318.3791317}
\bibfield{author}{\bibinfo{person}{Mingming Li}, \bibinfo{person}{Dingning
  Cao}, \bibinfo{person}{Xiang Chang}, \bibinfo{person}{Karla Sahin},
  \bibinfo{person}{Stefanie Mueller}, {and} \bibinfo{person}{Jiaji Li}.}
  \bibinfo{year}{2026}\natexlab{a}.
\newblock \showarticletitle{Xspine: Integrating Motion Sensing Capability into
  Dynamic Structures Using Multi-material FDM 3D Printing}. In
  \bibinfo{booktitle}{\emph{Proceedings of the 2026 CHI Conference on Human
  Factors in Computing Systems}} \emph{(\bibinfo{series}{CHI '26})}.
  \bibinfo{publisher}{Association for Computing Machinery},
  \bibinfo{address}{New York, NY, USA}, Article \bibinfo{articleno}{1351},
  \bibinfo{numpages}{18}~pages.
\newblock
\showISBNx{9798400722783}
\href{https://doi.org/10.1145/3772318.3791317}{doi:\nolinkurl{10.1145/3772318.3791317}}


\bibitem[Li et~al\mbox{.}(2025b)]%
        {10.1145/3746058.3758350}
\bibfield{author}{\bibinfo{person}{Mingming Li}, \bibinfo{person}{Jiaji Li},
  \bibinfo{person}{Haotian Chen}, \bibinfo{person}{Dingning Cao},
  \bibinfo{person}{Karla Sahin}, {and} \bibinfo{person}{Stefanie Mueller}.}
  \bibinfo{year}{2025}\natexlab{b}.
\newblock \showarticletitle{Integrating Motion Sensing into 3D-Printed Bending
  Structures}. In \bibinfo{booktitle}{\emph{Adjunct Proceedings of the 38th
  Annual ACM Symposium on User Interface Software and Technology}}
  \emph{(\bibinfo{series}{UIST Adjunct '25})}. \bibinfo{publisher}{Association
  for Computing Machinery}, \bibinfo{address}{New York, NY, USA}, Article
  \bibinfo{articleno}{148}, \bibinfo{numpages}{3}~pages.
\newblock
\showISBNx{9798400720369}
\href{https://doi.org/10.1145/3746058.3758350}{doi:\nolinkurl{10.1145/3746058.3758350}}


\bibitem[Lin et~al\mbox{.}(2022)]%
        {10.1145/3491102.3502113}
\bibfield{author}{\bibinfo{person}{Hongnan Lin}, \bibinfo{person}{Liang He},
  \bibinfo{person}{Fangli Song}, \bibinfo{person}{Yifan Li},
  \bibinfo{person}{Tingyu Cheng}, \bibinfo{person}{Clement Zheng},
  \bibinfo{person}{Wei Wang}, {and} \bibinfo{person}{HyunJoo Oh}.}
  \bibinfo{year}{2022}\natexlab{}.
\newblock \showarticletitle{FlexHaptics: A Design Method for Passive Haptic
  Inputs Using Planar Compliant Structures}. In
  \bibinfo{booktitle}{\emph{Proceedings of the 2022 CHI Conference on Human
  Factors in Computing Systems}} (New Orleans, LA, USA)
  \emph{(\bibinfo{series}{CHI '22})}. \bibinfo{publisher}{Association for
  Computing Machinery}, \bibinfo{address}{New York, NY, USA}, Article
  \bibinfo{articleno}{169}, \bibinfo{numpages}{13}~pages.
\newblock
\showISBNx{9781450391573}
\href{https://doi.org/10.1145/3491102.3502113}{doi:\nolinkurl{10.1145/3491102.3502113}}


\bibitem[Lyytinen and Yoo(2002)]%
        {lyytinen2002ubiquitous}
\bibfield{author}{\bibinfo{person}{Kalle Lyytinen} {and}
  \bibinfo{person}{Youngjin Yoo}.} \bibinfo{year}{2002}\natexlab{}.
\newblock \showarticletitle{Ubiquitous computing}.
\newblock \bibinfo{journal}{\emph{Commun. ACM}} \bibinfo{volume}{45},
  \bibinfo{number}{12} (\bibinfo{year}{2002}), \bibinfo{pages}{63--65}.
\newblock


\bibitem[Megaro et~al\mbox{.}(2017)]%
        {10.1145/3072959.3073636}
\bibfield{author}{\bibinfo{person}{Vittorio Megaro}, \bibinfo{person}{Jonas
  Zehnder}, \bibinfo{person}{Moritz B\"{a}cher}, \bibinfo{person}{Stelian
  Coros}, \bibinfo{person}{Markus Gross}, {and} \bibinfo{person}{Bernhard
  Thomaszewski}.} \bibinfo{year}{2017}\natexlab{}.
\newblock \showarticletitle{A computational design tool for compliant
  mechanisms}.
\newblock \bibinfo{journal}{\emph{ACM Trans. Graph.}} \bibinfo{volume}{36},
  \bibinfo{number}{4}, Article \bibinfo{articleno}{82} (\bibinfo{date}{July}
  \bibinfo{year}{2017}), \bibinfo{numpages}{12}~pages.
\newblock
\showISSN{0730-0301}
\href{https://doi.org/10.1145/3072959.3073636}{doi:\nolinkurl{10.1145/3072959.3073636}}


\bibitem[Nakamaru et~al\mbox{.}(2017)]%
        {10.1145/3126594.3126666}
\bibfield{author}{\bibinfo{person}{Satoshi Nakamaru}, \bibinfo{person}{Ryosuke
  Nakayama}, \bibinfo{person}{Ryuma Niiyama}, {and} \bibinfo{person}{Yasuaki
  Kakehi}.} \bibinfo{year}{2017}\natexlab{}.
\newblock \showarticletitle{FoamSense: Design of Three Dimensional Soft Sensors
  with Porous Materials}. In \bibinfo{booktitle}{\emph{Proceedings of the 30th
  Annual ACM Symposium on User Interface Software and Technology}} (Qu\'{e}bec
  City, QC, Canada) \emph{(\bibinfo{series}{UIST '17})}.
  \bibinfo{publisher}{Association for Computing Machinery},
  \bibinfo{address}{New York, NY, USA}, \bibinfo{pages}{437–447}.
\newblock
\showISBNx{9781450349819}
\href{https://doi.org/10.1145/3126594.3126666}{doi:\nolinkurl{10.1145/3126594.3126666}}


\bibitem[Nguyen et~al\mbox{.}(2014)]%
        {10.1145/2632048.2636092}
\bibfield{author}{\bibinfo{person}{Vinh~P. Nguyen}, \bibinfo{person}{Sang~Ho
  Yoon}, \bibinfo{person}{Ansh Verma}, {and} \bibinfo{person}{Karthik Ramani}.}
  \bibinfo{year}{2014}\natexlab{}.
\newblock \showarticletitle{BendID: flexible interface for localized
  deformation recognition}. In \bibinfo{booktitle}{\emph{Proceedings of the
  2014 ACM International Joint Conference on Pervasive and Ubiquitous
  Computing}} (Seattle, Washington) \emph{(\bibinfo{series}{UbiComp '14})}.
  \bibinfo{publisher}{Association for Computing Machinery},
  \bibinfo{address}{New York, NY, USA}, \bibinfo{pages}{553–557}.
\newblock
\showISBNx{9781450329682}
\href{https://doi.org/10.1145/2632048.2636092}{doi:\nolinkurl{10.1145/2632048.2636092}}


\bibitem[Parzer et~al\mbox{.}(2017)]%
        {10.1145/3126594.3126652}
\bibfield{author}{\bibinfo{person}{Patrick Parzer}, \bibinfo{person}{Adwait
  Sharma}, \bibinfo{person}{Anita Vogl}, \bibinfo{person}{J\"{u}rgen Steimle},
  \bibinfo{person}{Alex Olwal}, {and} \bibinfo{person}{Michael Haller}.}
  \bibinfo{year}{2017}\natexlab{}.
\newblock \showarticletitle{SmartSleeve: Real-time Sensing of Surface and
  Deformation Gestures on Flexible, Interactive Textiles, using a Hybrid
  Gesture Detection Pipeline}. In \bibinfo{booktitle}{\emph{Proceedings of the
  30th Annual ACM Symposium on User Interface Software and Technology}}
  (Qu\'{e}bec City, QC, Canada) \emph{(\bibinfo{series}{UIST '17})}.
  \bibinfo{publisher}{Association for Computing Machinery},
  \bibinfo{address}{New York, NY, USA}, \bibinfo{pages}{565–577}.
\newblock
\showISBNx{9781450349819}
\href{https://doi.org/10.1145/3126594.3126652}{doi:\nolinkurl{10.1145/3126594.3126652}}


\bibitem[Pointner et~al\mbox{.}(2025)]%
        {10.1145/3746059.3747733}
\bibfield{author}{\bibinfo{person}{Andreas Pointner}, \bibinfo{person}{Thomas
  Preindl}, \bibinfo{person}{Mira~A. Haberfellner}, \bibinfo{person}{Nitzan
  Cohen}, \bibinfo{person}{Niko M\"{u}nzenrieder}, {and}
  \bibinfo{person}{Michael Haller}.} \bibinfo{year}{2025}\natexlab{}.
\newblock \showarticletitle{Embroidering Resonant Circuits for Inductive
  Pressure Sensing}. In \bibinfo{booktitle}{\emph{Proceedings of the 38th
  Annual ACM Symposium on User Interface Software and Technology}}
  \emph{(\bibinfo{series}{UIST '25})}. \bibinfo{publisher}{Association for
  Computing Machinery}, \bibinfo{address}{New York, NY, USA}, Article
  \bibinfo{articleno}{2}, \bibinfo{numpages}{7}~pages.
\newblock
\showISBNx{9798400720376}
\href{https://doi.org/10.1145/3746059.3747733}{doi:\nolinkurl{10.1145/3746059.3747733}}


\bibitem[Rendl et~al\mbox{.}(2014)]%
        {10.1145/2642918.2647405}
\bibfield{author}{\bibinfo{person}{Christian Rendl}, \bibinfo{person}{David
  Kim}, \bibinfo{person}{Sean Fanello}, \bibinfo{person}{Patrick Parzer},
  \bibinfo{person}{Christoph Rhemann}, \bibinfo{person}{Jonathan Taylor},
  \bibinfo{person}{Martin Zirkl}, \bibinfo{person}{Gregor Scheipl},
  \bibinfo{person}{Thomas Rothl\"{a}nder}, \bibinfo{person}{Michael Haller},
  {and} \bibinfo{person}{Shahram Izadi}.} \bibinfo{year}{2014}\natexlab{}.
\newblock \showarticletitle{FlexSense: a transparent self-sensing deformable
  surface}. In \bibinfo{booktitle}{\emph{Proceedings of the 27th Annual ACM
  Symposium on User Interface Software and Technology}} (Honolulu, Hawaii, USA)
  \emph{(\bibinfo{series}{UIST '14})}. \bibinfo{publisher}{Association for
  Computing Machinery}, \bibinfo{address}{New York, NY, USA},
  \bibinfo{pages}{129–138}.
\newblock
\showISBNx{9781450330695}
\href{https://doi.org/10.1145/2642918.2647405}{doi:\nolinkurl{10.1145/2642918.2647405}}


\bibitem[Sakura et~al\mbox{.}(2023)]%
        {10.1145/3623263.3623361}
\bibfield{author}{\bibinfo{person}{Rei Sakura}, \bibinfo{person}{Changyo Han},
  \bibinfo{person}{Yahui Lyu}, \bibinfo{person}{Keisuke Watanabe},
  \bibinfo{person}{Ryosuke Yamamura}, {and} \bibinfo{person}{Yasuaki Kakehi}.}
  \bibinfo{year}{2023}\natexlab{}.
\newblock \showarticletitle{LattiSense: A 3D-Printable Resistive Deformation
  Sensor with Lattice Structures}. In \bibinfo{booktitle}{\emph{Proceedings of
  the 8th ACM Symposium on Computational Fabrication}} (New York City, NY, USA)
  \emph{(\bibinfo{series}{SCF '23})}. \bibinfo{publisher}{Association for
  Computing Machinery}, \bibinfo{address}{New York, NY, USA}, Article
  \bibinfo{articleno}{2}, \bibinfo{numpages}{14}~pages.
\newblock
\showISBNx{9798400703195}
\href{https://doi.org/10.1145/3623263.3623361}{doi:\nolinkurl{10.1145/3623263.3623361}}


\bibitem[Sakura and Kakehi(2025)]%
        {Single-Stroke}
\bibfield{author}{\bibinfo{person}{Rei Sakura} {and} \bibinfo{person}{Yasuaki
  Kakehi}.} \bibinfo{year}{2025}\natexlab{}.
\newblock \showarticletitle{A 3D-Printed Touch Sensor with a Single-Stroke
  Conductive Path}. In \bibinfo{booktitle}{\emph{Proceedings of the Extended
  Abstracts of the CHI Conference on Human Factors in Computing Systems}}
  \emph{(\bibinfo{series}{CHI EA '25})}. \bibinfo{publisher}{Association for
  Computing Machinery}, \bibinfo{address}{New York, NY, USA}, Article
  \bibinfo{articleno}{38}, \bibinfo{numpages}{6}~pages.
\newblock
\showISBNx{9798400713958}
\href{https://doi.org/10.1145/3706599.3720023}{doi:\nolinkurl{10.1145/3706599.3720023}}


\bibitem[Schmitz et~al\mbox{.}(2015)]%
        {Capricate}
\bibfield{author}{\bibinfo{person}{Martin Schmitz},
  \bibinfo{person}{Mohammadreza Khalilbeigi}, \bibinfo{person}{Matthias
  Balwierz}, \bibinfo{person}{Roman Lissermann}, \bibinfo{person}{Max
  M\"{u}hlh\"{a}user}, {and} \bibinfo{person}{J\"{u}rgen Steimle}.}
  \bibinfo{year}{2015}\natexlab{}.
\newblock \showarticletitle{Capricate: A Fabrication Pipeline to Design and 3D
  Print Capacitive Touch Sensors for Interactive Objects}. In
  \bibinfo{booktitle}{\emph{Proceedings of the 28th Annual ACM Symposium on
  User Interface Software \& Technology}} (Charlotte, NC, USA)
  \emph{(\bibinfo{series}{UIST '15})}. \bibinfo{publisher}{Association for
  Computing Machinery}, \bibinfo{address}{New York, NY, USA},
  \bibinfo{pages}{253–258}.
\newblock
\showISBNx{9781450337793}
\href{https://doi.org/10.1145/2807442.2807503}{doi:\nolinkurl{10.1145/2807442.2807503}}


\bibitem[Schmitz et~al\mbox{.}(2017)]%
        {10.1145/3025453.3025663}
\bibfield{author}{\bibinfo{person}{Martin Schmitz}, \bibinfo{person}{J\"{u}rgen
  Steimle}, \bibinfo{person}{Jochen Huber}, \bibinfo{person}{Niloofar Dezfuli},
  {and} \bibinfo{person}{Max M\"{u}hlh\"{a}user}.}
  \bibinfo{year}{2017}\natexlab{}.
\newblock \showarticletitle{Flexibles: Deformation-Aware 3D-Printed Tangibles
  for Capacitive Touchscreens}. In \bibinfo{booktitle}{\emph{Proceedings of the
  2017 CHI Conference on Human Factors in Computing Systems}} (Denver,
  Colorado, USA) \emph{(\bibinfo{series}{CHI '17})}.
  \bibinfo{publisher}{Association for Computing Machinery},
  \bibinfo{address}{New York, NY, USA}, \bibinfo{pages}{1001–1014}.
\newblock
\showISBNx{9781450346559}
\href{https://doi.org/10.1145/3025453.3025663}{doi:\nolinkurl{10.1145/3025453.3025663}}


\bibitem[Sun et~al\mbox{.}(2022)]%
        {x-bridges_2022}
\bibfield{author}{\bibinfo{person}{Lingyun Sun}, \bibinfo{person}{Jiaji Li},
  \bibinfo{person}{Junzhe Ji}, \bibinfo{person}{Deying Pan},
  \bibinfo{person}{Mingming Li}, \bibinfo{person}{Kuangqi Zhu},
  \bibinfo{person}{Yitao Fan}, \bibinfo{person}{Yue Yang}, \bibinfo{person}{Ye
  Tao}, {and} \bibinfo{person}{Guanyun Wang}.} \bibinfo{year}{2022}\natexlab{}.
\newblock \showarticletitle{X-{Bridges}: {Designing} {Tunable} {Bridges} to
  {Enrich} {3D} {Printed} {Objects}' {Deformation} and {Stiffness}}. In
  \bibinfo{booktitle}{\emph{Proceedings of the 35th {Annual} {ACM} {Symposium}
  on {User} {Interface} {Software} and {Technology}}}.
  \bibinfo{publisher}{ACM}, \bibinfo{address}{Bend OR USA},
  \bibinfo{pages}{1--12}.
\newblock
\showISBNx{978-1-4503-9320-1}
\href{https://doi.org/10.1145/3526113.3545710}{doi:\nolinkurl{10.1145/3526113.3545710}}


\bibitem[Tejada et~al\mbox{.}(2020)]%
        {10.1145/3313831.3376136}
\bibfield{author}{\bibinfo{person}{Carlos~E. Tejada}, \bibinfo{person}{Raf
  Ramakers}, \bibinfo{person}{Sebastian Boring}, {and} \bibinfo{person}{Daniel
  Ashbrook}.} \bibinfo{year}{2020}\natexlab{}.
\newblock \showarticletitle{AirTouch: 3D-printed Touch-Sensitive Objects Using
  Pneumatic Sensing}. In \bibinfo{booktitle}{\emph{Proceedings of the 2020 CHI
  Conference on Human Factors in Computing Systems}} (Honolulu, HI, USA)
  \emph{(\bibinfo{series}{CHI '20})}. \bibinfo{publisher}{Association for
  Computing Machinery}, \bibinfo{address}{New York, NY, USA},
  \bibinfo{pages}{1–10}.
\newblock
\showISBNx{9781450367080}
\href{https://doi.org/10.1145/3313831.3376136}{doi:\nolinkurl{10.1145/3313831.3376136}}


\bibitem[Wang et~al\mbox{.}(2024)]%
        {X-hair}
\bibfield{author}{\bibinfo{person}{Guanyun Wang}, \bibinfo{person}{Junzhe Ji},
  \bibinfo{person}{Yunkai Xu}, \bibinfo{person}{Lei Ren},
  \bibinfo{person}{Xiaoyang Wu}, \bibinfo{person}{Chunyuan Zheng},
  \bibinfo{person}{Xiaojing Zhou}, \bibinfo{person}{Xin Tang},
  \bibinfo{person}{Boyu Feng}, \bibinfo{person}{Lingyun Sun},
  \bibinfo{person}{Ye Tao}, {and} \bibinfo{person}{Jiaji Li}.}
  \bibinfo{year}{2024}\natexlab{}.
\newblock \showarticletitle{X-Hair: 3D Printing Hair-like Structures with
  Multi-form, Multi-property and Multi-function}. In
  \bibinfo{booktitle}{\emph{Proceedings of the 37th Annual ACM Symposium on
  User Interface Software and Technology}} (Pittsburgh, PA, USA)
  \emph{(\bibinfo{series}{UIST '24})}. \bibinfo{publisher}{Association for
  Computing Machinery}, \bibinfo{address}{New York, NY, USA}, Article
  \bibinfo{articleno}{65}, \bibinfo{numpages}{14}~pages.
\newblock
\showISBNx{9798400706288}
\href{https://doi.org/10.1145/3654777.3676360}{doi:\nolinkurl{10.1145/3654777.3676360}}


\bibitem[Watanabe et~al\mbox{.}(2021)]%
        {10.1145/3411763.3451547}
\bibfield{author}{\bibinfo{person}{Keisuke Watanabe}, \bibinfo{person}{Ryosuke
  Yamamura}, {and} \bibinfo{person}{Yasuaki Kakehi}.}
  \bibinfo{year}{2021}\natexlab{}.
\newblock \showarticletitle{foamin: A Deformable Sensor for Multimodal Inputs
  Based on Conductive Foam with a Single Wire}. In
  \bibinfo{booktitle}{\emph{Extended Abstracts of the 2021 CHI Conference on
  Human Factors in Computing Systems}} (Yokohama, Japan)
  \emph{(\bibinfo{series}{CHI EA '21})}. \bibinfo{publisher}{Association for
  Computing Machinery}, \bibinfo{address}{New York, NY, USA}, Article
  \bibinfo{articleno}{189}, \bibinfo{numpages}{4}~pages.
\newblock
\showISBNx{9781450380959}
\href{https://doi.org/10.1145/3411763.3451547}{doi:\nolinkurl{10.1145/3411763.3451547}}


\bibitem[Yang et~al\mbox{.}(2025)]%
        {10.1145/3706598.3714307}
\bibfield{author}{\bibinfo{person}{Humphrey Yang}, \bibinfo{person}{I-Chao
  Shen}, \bibinfo{person}{Nikolas Martelaro}, \bibinfo{person}{Bo Zhu},
  \bibinfo{person}{Haoran Xie}, \bibinfo{person}{Takeo Igarashi}, {and}
  \bibinfo{person}{Lining Yao}.} \bibinfo{year}{2025}\natexlab{}.
\newblock \showarticletitle{CompAct: Designing Interconnected Compliant
  Mechanisms with Targeted Actuation Transmissions}. In
  \bibinfo{booktitle}{\emph{Proceedings of the 2025 CHI Conference on Human
  Factors in Computing Systems}} \emph{(\bibinfo{series}{CHI '25})}.
  \bibinfo{publisher}{Association for Computing Machinery},
  \bibinfo{address}{New York, NY, USA}, Article \bibinfo{articleno}{1},
  \bibinfo{numpages}{19}~pages.
\newblock
\showISBNx{9798400713941}
\href{https://doi.org/10.1145/3706598.3714307}{doi:\nolinkurl{10.1145/3706598.3714307}}


\bibitem[Yoon et~al\mbox{.}(2017)]%
        {10.1145/3126594.3126654}
\bibfield{author}{\bibinfo{person}{Sang~Ho Yoon}, \bibinfo{person}{Ke Huo},
  \bibinfo{person}{Yunbo Zhang}, \bibinfo{person}{Guiming Chen},
  \bibinfo{person}{Luis Paredes}, \bibinfo{person}{Subramanian Chidambaram},
  {and} \bibinfo{person}{Karthik Ramani}.} \bibinfo{year}{2017}\natexlab{}.
\newblock \showarticletitle{iSoft: A Customizable Soft Sensor with Real-time
  Continuous Contact and Stretching Sensing}. In
  \bibinfo{booktitle}{\emph{Proceedings of the 30th Annual ACM Symposium on
  User Interface Software and Technology}} (Qu\'{e}bec City, QC, Canada)
  \emph{(\bibinfo{series}{UIST '17})}. \bibinfo{publisher}{Association for
  Computing Machinery}, \bibinfo{address}{New York, NY, USA},
  \bibinfo{pages}{665–678}.
\newblock
\showISBNx{9781450349819}
\href{https://doi.org/10.1145/3126594.3126654}{doi:\nolinkurl{10.1145/3126594.3126654}}


\bibitem[Zhou et~al\mbox{.}(2023)]%
        {IMU}
\bibfield{author}{\bibinfo{person}{Zihong Zhou}, \bibinfo{person}{Pei Chen},
  \bibinfo{person}{Yinyu Lu}, \bibinfo{person}{Qiang Cui},
  \bibinfo{person}{Deying Pan}, \bibinfo{person}{Yilun Liu},
  \bibinfo{person}{Jiaji Li}, \bibinfo{person}{Yang Zhang}, \bibinfo{person}{Ye
  Tao}, \bibinfo{person}{Xuanhui Liu}, \bibinfo{person}{Lingyun Sun}, {and}
  \bibinfo{person}{Guanyun Wang}.} \bibinfo{year}{2023}\natexlab{}.
\newblock \showarticletitle{3D Deformation Capture via a Configurable
  Self-Sensing IMU Sensor Network}.
\newblock \bibinfo{journal}{\emph{Proc. ACM Interact. Mob. Wearable Ubiquitous
  Technol.}} \bibinfo{volume}{7}, \bibinfo{number}{1}, Article
  \bibinfo{articleno}{42} (\bibinfo{date}{March} \bibinfo{year}{2023}),
  \bibinfo{numpages}{24}~pages.
\newblock
\href{https://doi.org/10.1145/3580874}{doi:\nolinkurl{10.1145/3580874}}


\end{thebibliography}

\color{black}
\appendix

\section{APPENDIX}

\subsection{Printing Setting for 3D Printer}\label{print}

\begin{figure}[h]
    \includegraphics[width=1\linewidth]{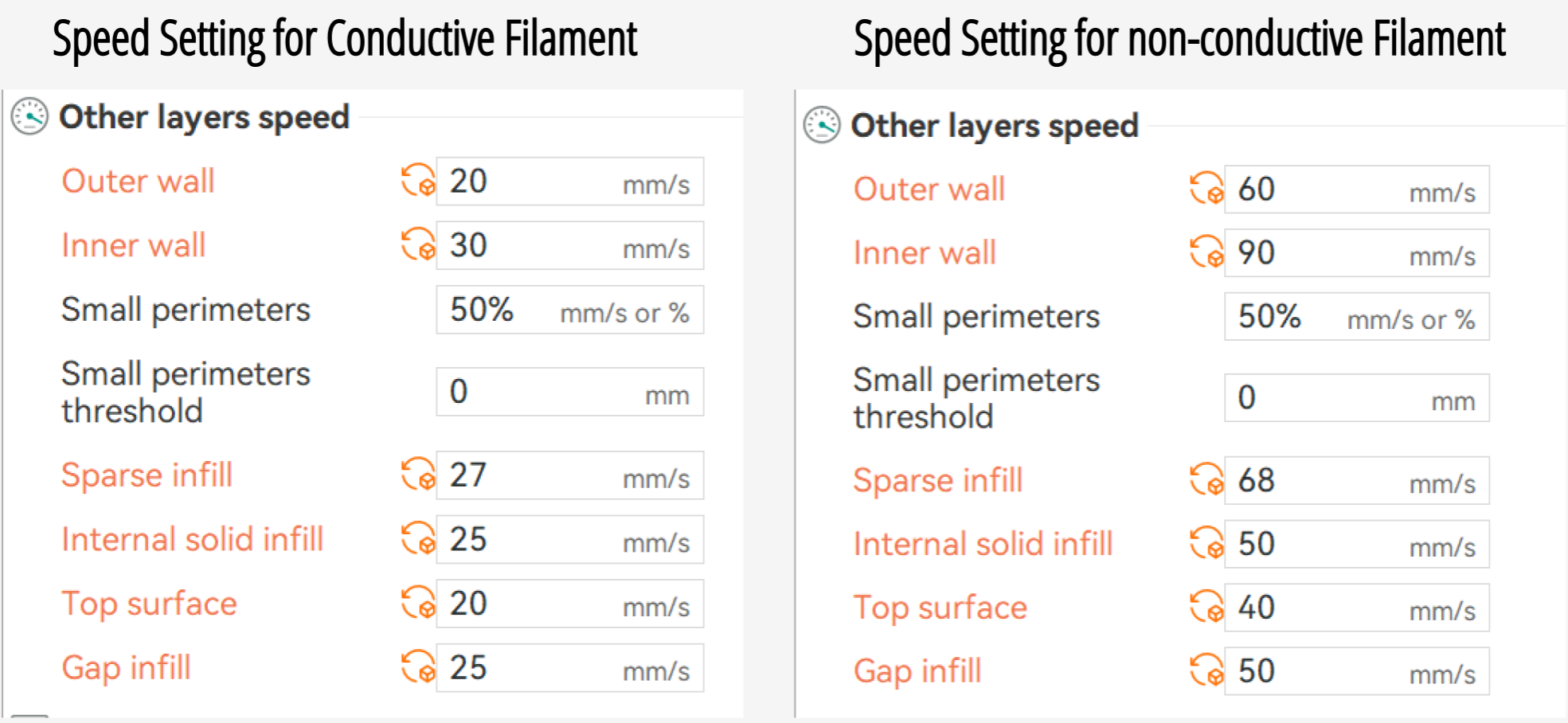}
    \caption{Example printing parameters for fabricating the X-Hinges using the multi-material FDM printer, including the Prusa XL, Bambu H2D, and Snapmaker U1}
    \Description{Side-by-side slicer speed settings for conductive and nonconductive filament. Conductive-filament speeds are: outer wall 20 mm/s, inner wall 30 mm/s, small perimeters 50 percent with 0 mm threshold, sparse infill 27 mm/s, internal solid infill 25 mm/s, top surface 20 mm/s, and gap infill 25 mm/s. Nonconductive-filament speeds are: outer wall 60 mm/s, inner wall 90 mm/s, small perimeters 50 percent with 0 mm threshold, sparse infill 68 mm/s, internal solid infill 50 mm/s, top surface 40 mm/s, and gap infill 50 mm/s.}
    \label{printing_parameter}
\end{figure}

\subsection{Mechanical Properties Evaluation}\label{evaluation_0}
\begin{figure}[h]
    \includegraphics[width=1\linewidth]{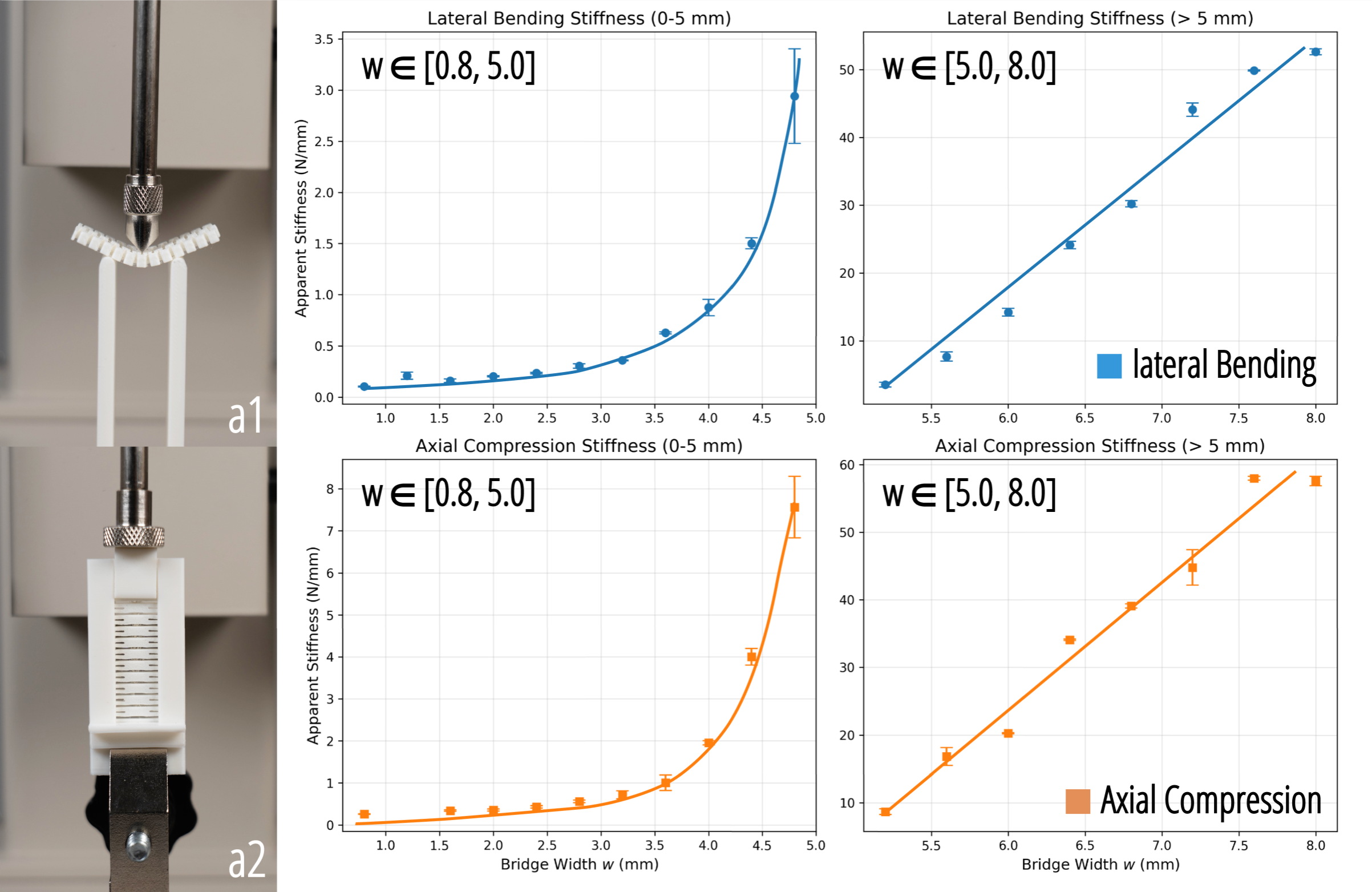}
    \caption{Mechanical evaluation of X-Hinges. (a) Experimental setup for lateral bending and axial compression tests. (b) Apparent stiffness as a function of bridge width $w$.}
    \Description{Mechanical stiffness evaluation. At left, photographs show a universal testing machine performing lateral three-point bending and axial compression on white X-Hinges specimens. Four plots show apparent stiffness versus bridge width w. From 0.8 to 5.0 mm, both lateral and axial stiffness increase nonlinearly, reaching about 3 and 7.5 N/mm, respectively. From 5.0 to 8.0 mm, the relation is approximately linear: lateral stiffness rises from about 2.5 to 52 N/mm, and axial stiffness from about 8.5 to 58 N/mm. Points include error bars and fitted trend lines.}
    \label{Evaluation1}
\end{figure}

We investigated the stiffness of X-Hinges across two primary motion directions (lateral bending and axial compression) and characterized how bridge width $w$ governs structural compliance.

We fabricated X-Hinges specimens with bridge widths $w$ ranging from \(0.8\,\mathrm{mm}\) to \(8.0\,\mathrm{mm}\) while keeping other dimensions fixed (cross-section\red{(L)} \(20\,\mathrm{mm}\), length \(40\,\mathrm{mm}\), thickness \(h = 3\,\mathrm{mm}\), spacing \(d = 0.8\,\mathrm{mm}\)). Lateral bending stiffness was measured using three-point bending, and axial stiffness through compression tests. Applied force and displacement were recorded at \(1\,\mathrm{mm}\) and \(2\,\mathrm{mm}\) deformations.

Results show that bridge width $w$ significantly governs stiffness with a biphasic relationship (Figure~\ref{Evaluation1}b). When $w < L/4$ (\(5\,\mathrm{mm}\)), \red{stiffness increased strongly and nonlinearly with $w$ due to torsional-bending coupling in slender bridges}. When $w \geq L/4$, scaling transitions to $k \propto w^{1}$ as the structure behaves like bulk material. This enables stiffness tuning across two orders of magnitude.

\subsection{Electromechanical Durability}\label{evaluation_2}

\begin{figure}[h]
    \includegraphics[width=1\linewidth]{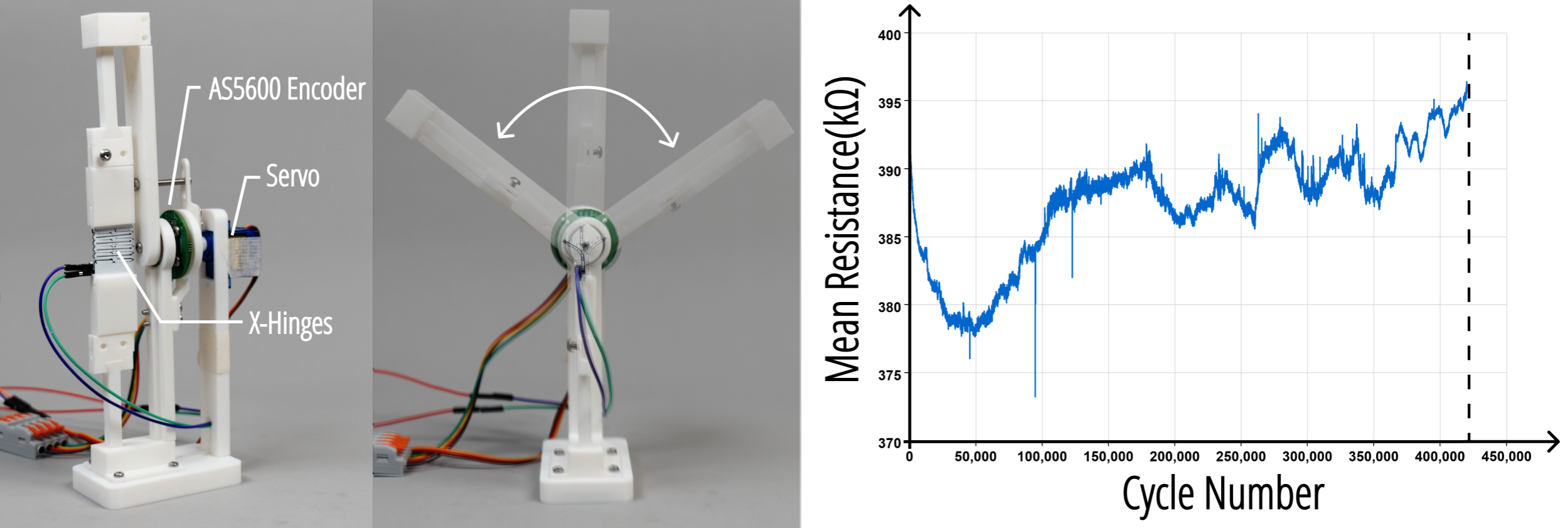}
    \caption{Electromechanical durability evaluation of X-Hinges. (a) Fatigue test setup. (b) Mean resistance over 420,000 cycles.}
    \Description{Electromechanical durability test. (a) Photographs show a servo-driven fixture with an AS5600 magnetic encoder repeatedly bending an X-Hinges specimen between two angular extremes. (b) Mean resistance over 420,000 cycles first falls from about 390 to 378 kiloohms, then recovers and fluctuates mainly between 386 and 395 kiloohms, ending near 396 kiloohms at the dashed 420,000-cycle marker. The specimen remained structurally intact after the test.}
    \label{Evaluation2}
\end{figure}

To evaluate the long-term reliability of X-Hinges, we designed a fatigue tester (Figure~\ref{Evaluation2}a) driven by a servo motor and instrumented with a magnetic angle encoder. Specimens with fixed geometry ($L = 20\,\mathrm{mm}$, $h = 1.2\,\mathrm{mm}$, $d = 0.8\,\mathrm{mm}$, $w = 1.2\,\mathrm{mm}$) were cycled through $[-90^\circ,\,80^\circ]$ at \(2\,\mathrm{s}\) per cycle. Resistance was recorded at \(50\,\mathrm{Hz}\) throughout the experiment.

After 420,000 cycles over 233 hours, the structure remained intact without visible fatigue or failure. The mean resistance remained stable within 375–395~k$\Omega$ (<5.5\% fluctuation, Figure~\ref{Evaluation2}b), demonstrating consistent sensing performance. For high-precision applications, we recommend periodic recalibration using the self-calibration baseline (Section~\ref{self-calibration}) to compensate for subtle baseline shifts.

\subsection{Hardware Performance Comparison with Voltage Divider Approach}
\label{hardware_comparison}

\begin{figure}[h]\centering
  \includegraphics[width=0.9\linewidth]{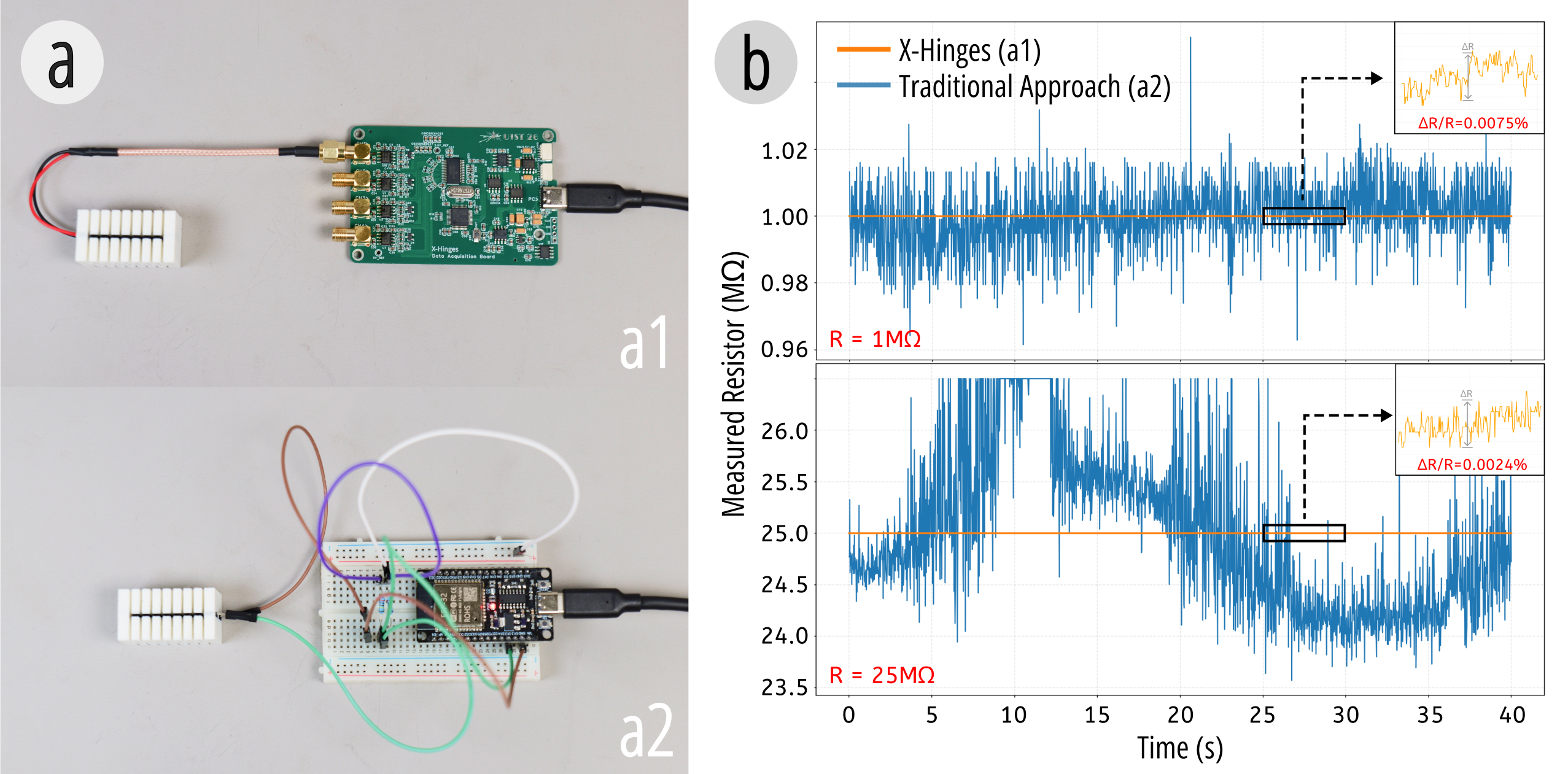}
  \caption{Hardware comparison. (a) Voltage divider with ESP32 ADC vs. constant-voltage excitation with transimpedance amplifier and ADS1256. (b) Resistance measurements over time (top: $1 M\Omega$; bottom: $25 M\Omega$). Insets show noise differences.}
  \Description{Hardware measurement comparison. (a1) A fixed resistor and small printed X-Hinge connect directly to the green X-Hinges acquisition board. (a2) The traditional setup connects a similar specimen through a breadboard voltage divider to an ESP32 board. (b) Resistance traces over 40 seconds compare X-Hinges in orange with the traditional approach in blue. For 1 megaohm, the orange trace stays nearly flat while the blue trace fluctuates visibly around 1 megaohm. For 25 megaohms, the orange trace again stays flat while the blue trace exhibits large noise and multi-megaohm drift. Insets magnify the orange variations, showing relative changes of 0.0075 percent and 0.0024 percent.}
  \label{compare}
\end{figure}

We compared our constant-voltage current-sensing approach with the traditional voltage divider method using fixed resistors at 50Hz sampling rate. Figure~\ref{compare} shows the hardware setups and measured resistance stability.

\begin{table}[h]
\caption{Performance comparison between voltage divider and X-Hinges hardware}
\Description{Performance comparison between a conventional voltage divider and X-Hinges resistance-acquisition hardware. For a 1 megaohm resistor, SNR improves by 43.75 dB, noise is reduced 160 times, divider resolution is 28.2 kiloohms, and X-Hinges resolution is 176 ohms. At 2 megaohms: 45.54 dB, 198 times, 60.5 kiloohms, and 306 ohms. At 6 megaohms: 38.16 dB, 87 times, 236 kiloohms, and 2.7 kiloohms. At 25 megaohms: 88.53 dB, 27,415 times, 21.9 megaohms, and 798 ohms. The largest improvement occurs at 25 megaohms.}
\label{tab:hardware_comparison}
\small
\begin{tabular}{lcccc}
\hline
\textbf{Resistor} & \textbf{SNR} & \textbf{Noise} & \textbf{Resolution} & \textbf{Resolution} \\
& \textbf{Improvement} & \textbf{Reduction} & \textbf{(Divider)} & \textbf{(X-Hinges)} \\
\hline
$1 M\Omega$  & +43.75 dB & 160$\times$    & 28.2 k$\Omega$ & 176 $\Omega$ \\
$2 M\Omega$  & +45.54 dB & 198$\times$    & 60.5 k$\Omega$ & 306 $\Omega$ \\
$6 M\Omega$  & +38.16 dB & 87$\times$     & 236 k$\Omega$  & 2.7 k$\Omega$ \\
$25 M\Omega$ & +88.53 dB & 27,415$\times$ & 21.9 M$\Omega$ & 798 $\Omega$ \\
\hline
\end{tabular}
\end{table}

Table~\ref{tab:hardware_comparison} summarizes the quantitative performance improvements. Our approach achieves 38-89 dB SNR improvement across all resistance ranges, with the most dramatic improvement at $25 M\Omega$ (27,415$\times$ noise reduction). The relative current-measurement resolution remains below 0.05\% across all tested resistance levels, enabling sub-microampere detection in high-resistance sensing elements. At $25 M\Omega$ (typical for conductive TPU), the overall performance improvement factor is $2.7\times 10^4$.

\subsection{Hardware Schematics}\label{hardware_sch}
\begin{figure}[h]
  \includegraphics[width=0.96\linewidth]{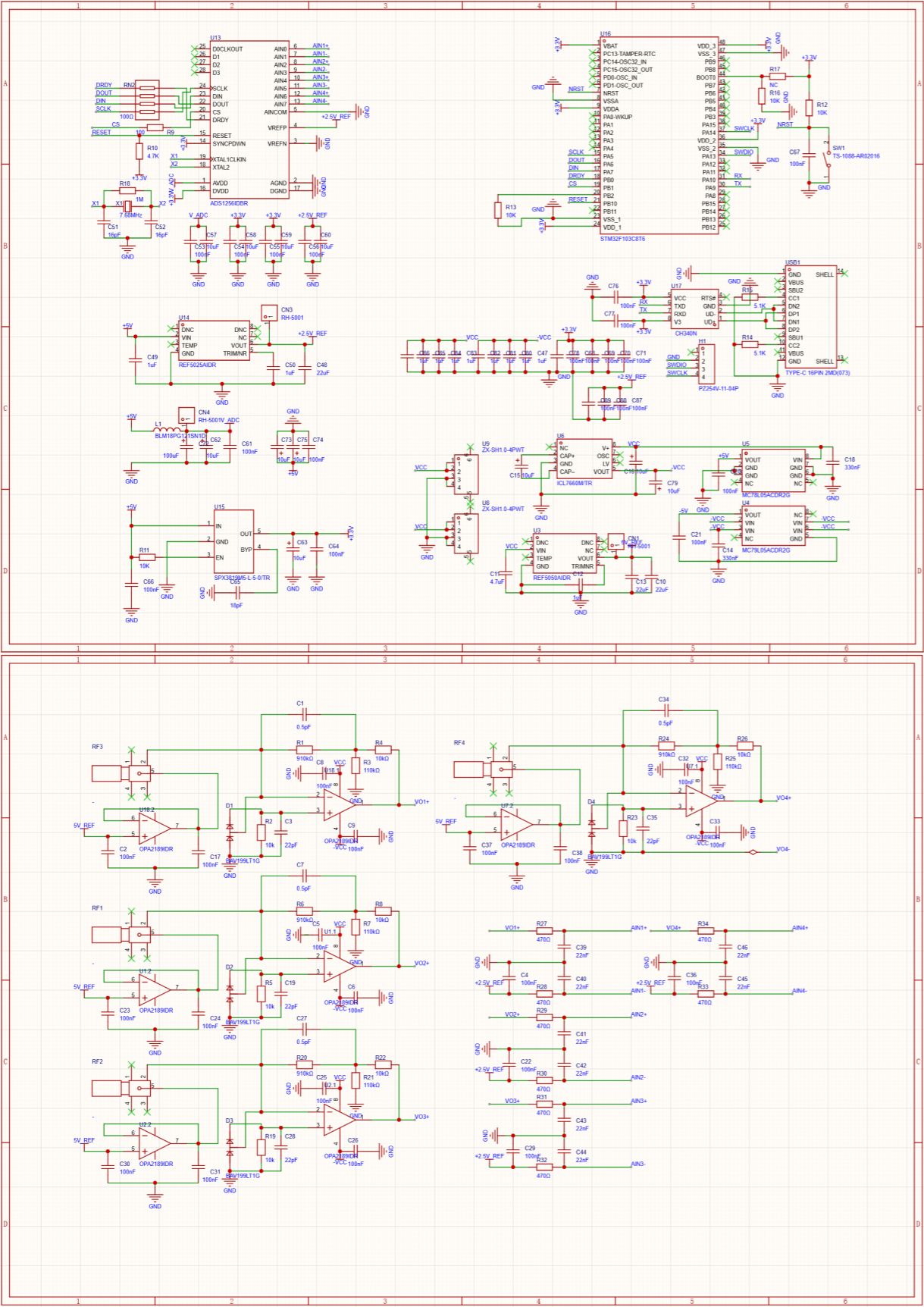}
  \caption{Sensing hardware: data acquisition circuit and sensing sampling circuit}
  \Description{Two-page circuit schematic for the X-Hinges sensing hardware. The upper acquisition schematic includes an ADS1256 analog-to-digital converter, STM32F103 microcontroller, CH340N USB interface, voltage reference, clock, USB-C connector, and regulated power rails. The lower sensing schematic contains four repeated input channels. Each channel provides stable 5 V reference excitation, protection and filtering, and an OPA2189-based transimpedance stage; the four conditioned differential outputs are filtered and routed to the ADS1256 analog inputs.}
  \label{sch}
\end{figure}

\red{Figure~\ref{sch} presents the complete schematics of the X-Hinges sensing hardware. The data acquisition circuit handles signal digitization, processing, and communication, while the repeated sensing channels provide stable voltage excitation and convert the resulting feedback currents into measurable voltages through transimpedance amplifiers.}

\subsection{Multi-DOF Decoding Details}\label{dof_decoding_demonstration}
\noindent\textbf{Network Architecture.}
The model takes a sliding window of $W = 256$ frames
(\(\approx 5.1\,\mathrm{s}\)) of $C$-channel resistance readings, augmented with
first-order temporal differences, yielding an input tensor of shape
$(W{-}1) \times 2C$. Four stacked temporal blocks with channel sizes
[64, 128, 128, 256], kernel size 7, and exponentially increasing dilation
rates ($d = 1, 2, 4, 8$) process the sequence. The output is aggregated via
adaptive average pooling and mapped to the target DOFs through a linear head.

\begin{figure}[h]
  \includegraphics[width=0.96\linewidth]{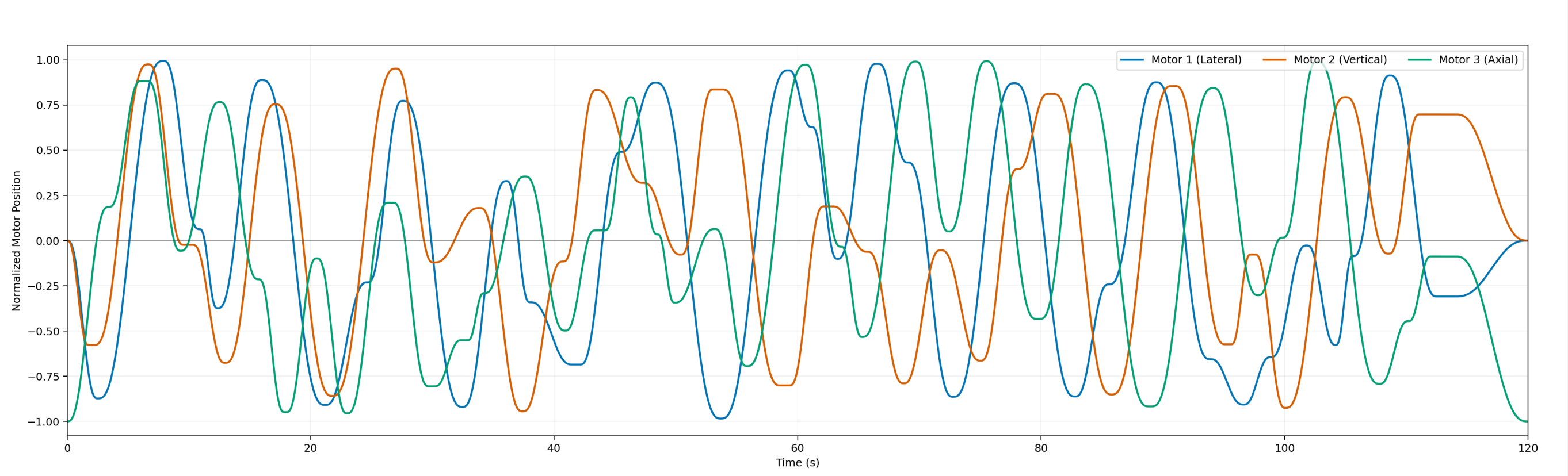}
  \caption{Randomized multi-DOF excitation for data collection.}
  \Description{Randomized multi-DOF excitation trajectories over 120 seconds. Three smooth, independently varying curves span normalized motor positions from -1 to +1: lateral motor in blue, vertical motor in orange, and axial motor in green. Their amplitudes, speeds, reversals, and occasional dwell intervals differ throughout the recording, producing overlapping but non-synchronized motions that cover combinations across the three-axis motion space.}
  \label{random_motion}
\end{figure}

\vspace{0.2cm}
\noindent\textbf{Data Collection.}
\red{For data collection, we used the actuation system described in
Section~\ref{evaluation_integrated} to simultaneously actuate all three
degrees of freedom of the X-Hinges. To increase the coverage of the coupled
motion space, the three axes followed independently randomized, smooth
command trajectories with varying amplitudes, speeds, and dwell times
(Figure~\ref{random_motion}). During actuation, we synchronously recorded
the resistance channels associated with the three sensing configurations
and the three motor-encoder signals. The resistance histories were used as
model inputs, while the encoder readings were converted into lateral angle,
vertical angle, and axial displacement as reference labels. In total, we
collected one continuous \(2621.84\,\mathrm{s}\) recording at \(50\,\mathrm{Hz}\), yielding
131,093 synchronized samples. The recording was divided chronologically
into training, validation, and test sets in a 70\%/15\%/15\% ratio before
temporal-window construction.}

\begin{figure}[h]
  \includegraphics[width=0.96\linewidth]{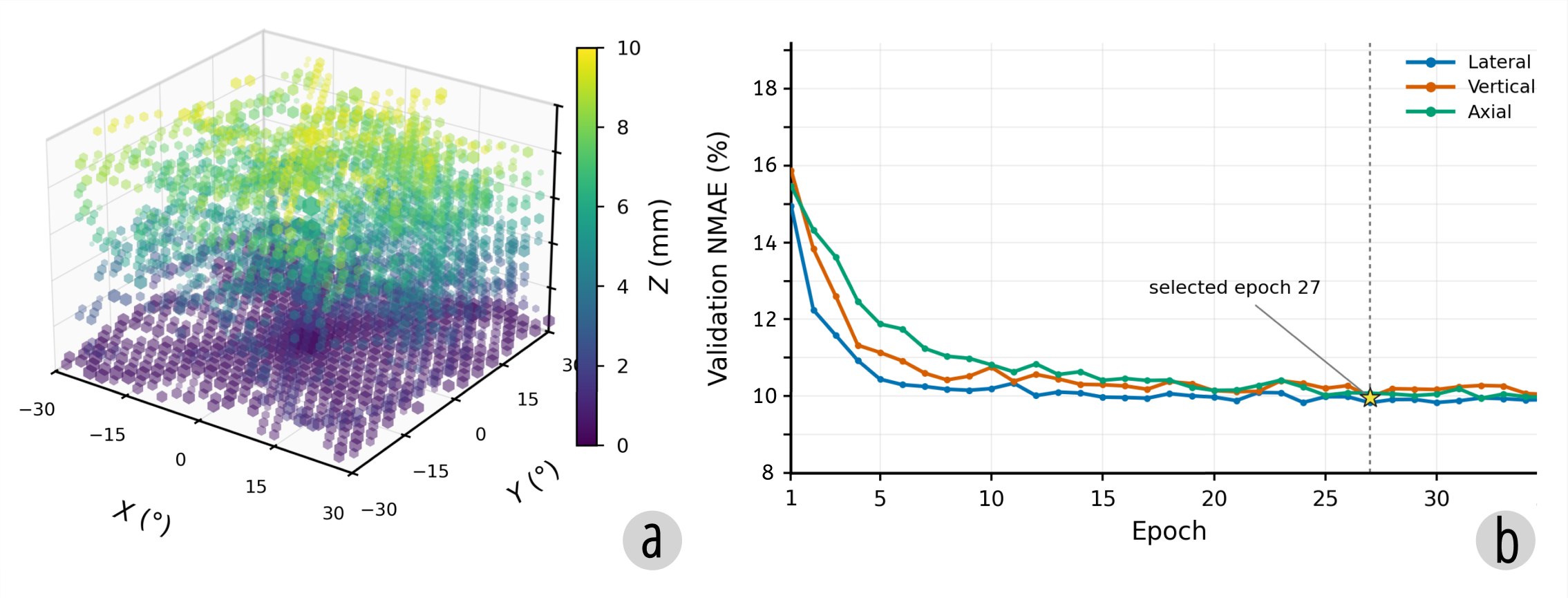}
  \caption{Training data and model convergence. (a) Training sample coverage across the motion space. (b) Validation NMAE over epochs.}
  \Description{Training-data coverage and model convergence. (a) A dense three-dimensional point cloud covers lateral X and vertical Y angles from approximately -30 to +30 degrees and axial Z displacement from 0 to 10 mm. Point color also encodes Z from purple at 0 to yellow at 10, showing coverage throughout the coupled motion volume. (b) Validation normalized mean absolute error for lateral, vertical, and axial outputs starts near 15-16 percent, drops rapidly to about 10-12 percent within five epochs, and then plateaus near 10 percent. A star and vertical dashed line mark epoch 27 as the selected checkpoint.}
  \label{data_training}
\end{figure}

\vspace{0.2cm}
\noindent\textbf{Training.}
The model is trained with AdamW ($\text{lr} = 10^{-3}$), Smooth L1 loss,
and early stopping. Input features and labels are standardized
independently. \red{Figure~\ref{data_training}a shows the distribution of
collected training samples across the three-DOF motion space, and
Figure~\ref{data_training}b shows the validation NMAE curves during
training, from which the checkpoint at epoch 27 was selected for
deployment.}

\vspace{0.2cm}
\noindent\textbf{Results.}
\red{We deploy the trained TCN on a three-DOF X-Hinges. The TCN achieves
mean absolute errors of $7.35^\circ$, $6.59^\circ$, and
\(1.35\,\mathrm{mm}\) for lateral angle, vertical angle, and axial
displacement, respectively.}

\color{black}

\end{document}